\RequirePackage{etoolbox}
\csdef{input@path}{{style/}{graphics/}}
\documentclass[MSNbibl,nameyear,dvips]{arxstspdf}
\makeatletter
   \@ifpackageloaded{graphicx}{}{\usepackage{graphicx}}
\makeatother

\usepackage{algorithm}
\usepackage{algorithmic}
\usepackage{graphicx}
\usepackage{url,breakurl}

\usepackage{flushend}
\usepackage{stfloats}

\volume{30}
\issue{1}
\pubyear{2015}
\firstpage{72}
\lastpage{95}
\doi{10.1214/14-STS498} 
\docsubty{FLA}

\makeatletter
\renewcommand{\epsilon}{\varepsilon}
\renewcommand{\citep}[1]{(\citeauthor{#1}, \citeyear{#1})}
\newproclaim{ass}{Assumption}
\newtheorem{them}{Theorem}
\newproclaim{rem}{Remark}
\newproclaim{res}{Result}
\newcommand{\vectr}[1]{\boldsymbol{#1}}
\newcommand{\vect}[1]{\mathbf{#1}}
\makeatother

\begin{document}
\begin{frontmatter}

\title{Bayesian Indirect Inference Using a~Parametric Auxiliary Model}%
\runtitle{Bayesian Indirect Inference}

\begin{aug}
\author[A]{\fnms{Christopher C.}~\snm{Drovandi}\corref{}\ead[label=e1]{c.drovandi@qut.edu.au}},
\author[A]{\fnms{Anthony N.}~\snm{Pettitt}\ead[label=e2]{a.pettitt@qut.edu.au}}
\and
\author[A]{\fnms{Anthony}~\snm{Lee}\ead[label=e3]{anthony.lee@warwick.ac.uk}}
\runauthor{C. C. Drovandi, A. N. Pettitt and A. Lee}

\affiliation{Queensland University of Technology, Queensland University
of Technology and University of Warwick}

\address[A]{Christopher C. Drovandi is Lecturer and Anthony N. Pettitt is
Professor, School of Mathematical Sciences, Queensland University of
Technology, Brisbane, Australia, 4000 \printead{e1,e2}.
Anthony Lee is Assistant Professor, Department of Statistics, University of Warwick,
Coventry, CV4 7AL, United Kingdom \printead{e3}.}
\end{aug}

%
\begin{abstract}
Indirect inference (II) is a methodology for estimating the parameters
of an intractable (generative) model on the basis of an alternative
parametric (auxiliary) model that is both analytically and
computationally easier to deal with. Such an approach has been well
explored in the classical literature but has received substantially
less attention in the Bayesian paradigm. The purpose of this paper is
to compare and contrast a collection of what we call parametric
Bayesian indirect inference (pBII) methods. One class of pBII methods
uses approximate Bayesian computation (referred to here as ABC II)
where the summary statistic is formed on the basis of the auxiliary
model, using ideas from II. Another approach proposed in the
literature, referred to here as parametric Bayesian indirect likelihood
(pBIL), uses the auxiliary likelihood as a replacement to the
intractable likelihood. We show that pBIL is a fundamentally different
approach to ABC II. We devise new theoretical results for pBIL to give
extra insights into its behaviour and also its differences with ABC II.
Furthermore, we examine in more detail the assumptions required to use
each pBII method. The results, insights and comparisons developed in
this paper are illustrated on simple examples and two other substantive
applications. The first of the substantive examples involves performing
inference for complex quantile distributions based on simulated data
while the second is for estimating the parameters of a trivariate
stochastic process describing the evolution of macroparasites within a
host based on real data. We create a novel framework called Bayesian
indirect likelihood (BIL) that encompasses pBII as well as general ABC
methods so that the connections between the methods can be established.
\end{abstract}

%

\end{frontmatter}

\section{Introduction}

Approximate Bayesian computation (ABC) now plays an important role in
performing (approximate) Bayesian inference for the parameter of a
proposed statistical model (called the generative model here) that has
an intractable likelihood. Despite the intense attention ABC has
recently received, the approach still suffers from several drawbacks.
An obvious disadvantage is the usual necessity to reduce the data to a
low dimensional summary statistic. This leads to a loss of information
that is difficult to quantify. The second, often less severe but
sometimes related, drawback is the computational challenge of achieving
stringent matching between the observed and simulated summary statistics.

In situations where an alternative parametric model (referred to as an
auxiliary model) can be formulated that has a tractable likelihood, the
methodology known as indirect inference (II) (see, e.g.,
Gourieroux, Monfort and
Renault, \citeyear{Gourieroux1993} and \cite{Heggland2004}) is applicable. II has been
thoroughly examined in the classical framework. Most methods differ in
the way that observed and simulated data are compared via the auxiliary
model. We expand on this later in the article. For the moment, we note
that some key references are \citet{Gourieroux1993},
\citet{Smith1993} and \citet{Gallant1996}.

\begin{figure*}

\includegraphics{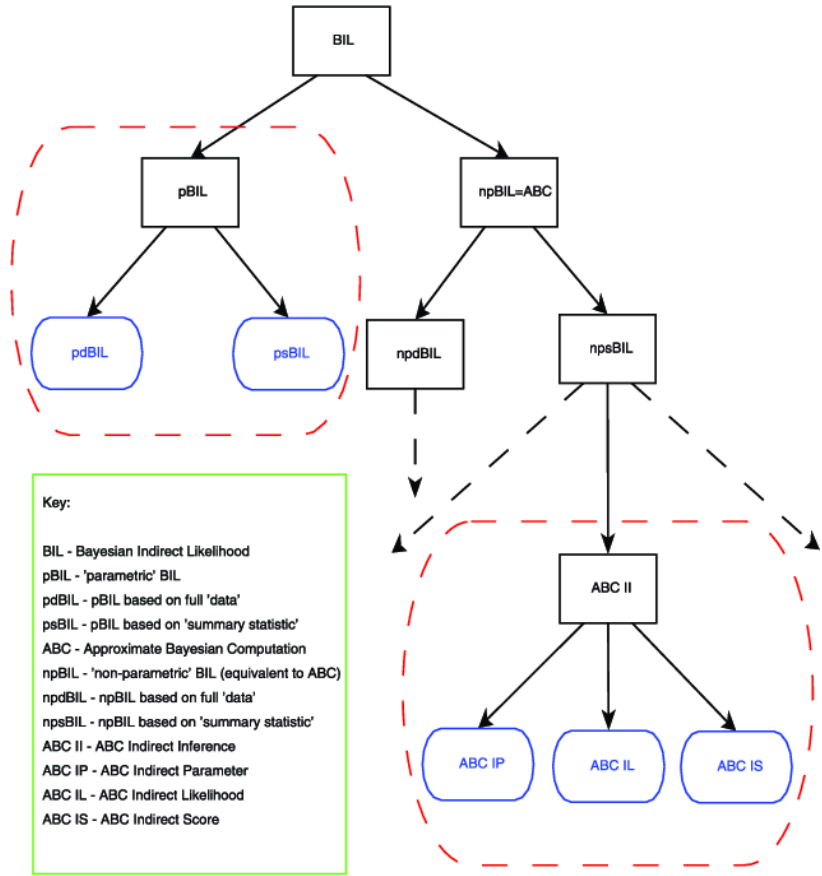}

\caption{The general BIL framework. Here, the rounded rectangles
indicate particular instances of methods. The dashed arrows indicate
that there are multiple instances of that class of method in the
literature that are not placed on this diagram. The dashed larger
rounded rectangles indicate the methods that this paper focusses on.
That is, BII methods that make use of, in some way, a parametric
auxiliary model (so-called pBII methods).}\label{fig:BIL}
\end{figure*}

However, II has been far less studied in the Bayesian paradigm.
\citet{DrovandiEtAl2011} developed an ABC approach that uses II to obtain
summary statistics. In particular, the estimated parameter of the
auxiliary model fitted to the data becomes the observed summary
statistic. We adopt a similar naming convention to \citet{Gleim} and
refer to this method as ABC IP where ``I'' stands for ``indirect'' and ``P''
stands for ``parameter''. \citet{Gleim} also list another method, ABC IL
(where ``L'' stands for ``likelihood''), which is essentially an ABC
version of \citet{Smith1993}. This approach follows ABC IP in the sense
that the parameter estimate of the auxiliary model is again the summary
statistic. However, the ABC discrepancy is based on the auxiliary
likelihood, rather than a direct comparison of the auxiliary parameters.

\citet{Gleim} advocate a slightly different approach to ABC with II,
which is effectively an ABC version of the classical approach in \citet
{Gallant1996}. Here, \citet{Gleim} use the score vector based on the
auxiliary model as the summary statistic, which is referred to as ABC
IS (here ``S'' stands for ``score''). The parameter value used in the score
is given by the MLE of the auxiliary model fitted to the observed data.
This approach can be far cheaper from a computational point of view
since it avoids an expensive fitting of the auxiliary model to each
dataset simulated from the generative model required in ABC IP and ABC IL.

Throughout the paper, the collection of approaches that use the
parametric auxiliary model to form summary statistics is referred to as
ABC II methods. An advantage of this approach over more traditional
summary statistics is that some insight can be gained on the potential
utility of the II summary statistic prior to the ABC analysis.
Additionally, if the auxiliary model is parsimonious, then the summary
statistic can be low-dimensional.

\citet{Gallant2009} (see also \cite{Reeves2005}) suggest an
alternative approach for combining II with Bayesian inference. This
method has similar steps to ABC IP and ABC IL but essentially uses the
likelihood of the auxiliary model as a replacement to the intractable
generative model likelihood. We note here that this is a fundamentally
different approach as it is not a standard ABC method. In particular,
there is no comparison of summary statistics and no need to choose an
ABC tolerance. Here, we refer to this method as parametric Bayesian
indirect likelihood (pBIL). The focus of this paper is the application
of a parametric auxiliary model for the full data, which we refer to as
pdBIL (where ``d'' stands for ``data''). However, the ideas in this paper
are equally applicable if a parametric model is applied at the summary
statistic (not necessarily obtained using ABC II techniques) level
(i.e., some data reduction technique has been applied; see \cite
{Blum2013}, for a review). This is referred to as psBIL (where ``s''
stands for ``summary statistic''). We show that the Bayesian version of
the synthetic likelihood method of \citet{Wood2010} is a psBIL method.
In the paper, we refer to the collection of ABC II and pBIL approaches
as pBII methods (``Bayesian'' ``Indirect'' ``Inference'' using a ``parametric''
auxiliary model).

In the process of reviewing these pBII methods, we create a novel
framework called Bayesian indirect likelihood (BIL) which encompasses
pBII as well as ABC methods generally. In particular, if a specific
nonparametric auxiliary model is selected (npBIL) instead of a
parametric one (pBIL), then the general ABC method is recovered. A
nonparametric kernel can be applied either at the full data (npdBIL)
or summary statistic (npsBIL) level. The ABC II approaches are thus a
special type of npsBIL method where the summary statistic is formed on
the basis of a parametric auxiliary model. This framework is shown in
Figure~\ref{fig:BIL}, which also highlights the methods that this paper
addresses.

This article does not develop any new algorithms for pBII. However,
this paper does make several interesting and useful contributions.
Firstly, we explore the pdBIL method in more detail theoretically, and
recognise that it is fundamentally different to ABC II. The behaviour
of this method is also substantially different. A technique sometimes
applied with classical II methods is to increase the simulated dataset
size beyond that of the observed data in order to reduce the
variability of the estimated quantities under the auxiliary model (see,
e.g., \cite{Smith1993}; \cite{Gourieroux1993}; \cite{Gallant2009}). We demonstrate
that pdBIL and ABC II behave differently for increasing size of the
simulated data. Our second contribution is to compare the assumptions
required for each pBII approach. Our theoretical and empirical results
\mbox{indicate} that the pBIL method will provide good approximations if the
auxiliary model is sufficiently flexible to estimate the likelihood of
the true model well based on parameter values within regions of
\mbox{non-negligible} posterior probability. ABC II methods rely on the
parameter estimate or the score of the auxiliary model to provide a
near-sufficient summary statistic. Finally, our creation of the general
BIL framework provides a clear way to see the connections between pBII
and other methods.

The paper is organised as follows. In Section~\ref{sec:notation}, the
notation used throughout the paper is defined. The ABC II methods are
reviewed in Section~\ref{sec:ABC II}. The pBIL approach is presented in
Section~\ref{sec:BIL}. The theoretical developments in this section,
which offer additional \mbox{insight} into the pBIL approximation, are new. In
addition, this section demonstrates how the synthetic likelihood
approach of \citet{Wood2010} is a pBIL method on the summary statistic
level. Section~\ref{sec:ABCasBIL} shows how ABC can be recovered as a
BIL method via a nonparametric choice of the auxiliary model.
Section~\ref{sec:compareBII} provides a comparison between ABC II and pdBIL.
The contributions of this article are demonstrated on examples with
varying complexity in Section~\ref{sec:Results}. The highlights of this
section include improved approximate inferences for quantile
distributions and a multivariate Markov jump process explaining the
evolution of parasites within a host. The article concludes with a
discussion in Section~\ref{sec:Discussion}.

\section{Notation} \label{sec:notation}

Consider an observed dataset $\vect{y}$ taking values in
$\mathsf{Y}$
of dimension $N$ assumed to have arisen from a generative model with an
intractable likelihood $p(\vect{y}|\vectr{\theta})$, where
$\vectr{\theta} \in\Theta$ is the parameter of this model.
Intractability here
refers to the inability to compute $p(\vect{y}|\vectr{\theta})$
pointwise as a function of $\vectr{\theta}$. We assume that there is a
second statistical model that has a tractable likelihood function. We
denote the likelihood of this auxiliary model by $p_A(\vect{y}|
\vectr{\phi})$, where $\vectr{\phi} \in\Phi$ denotes the parameter of this
auxiliary model. There does not necessarily need to be any obvious
connection between $\vectr{\theta}$ and $\vectr{\phi}$. The auxiliary
model could be purely a data analytic model that does not offer any
mechanistic explanation of how the observed data arose. The parameter
estimate of the auxiliary model when fitted to the observed data is
given by $\vectr{\phi}(\vect{y})$. Assuming a prior distribution on the
parameters of the generative model, $p(\vectr{\theta})$, our interest is
in sampling from the posterior distribution, $p(\vectr{\theta}|\vect
{y})$, or some approximation thereof.

We denote data simulated from the model as $\vect{x} \in\mathsf{Y}$.
In this paper, we also consider the effect of using $n$ independent
replicates of data simulated from the model, which we denote $\vect
{x}_{1:n} = (\vect{x}_1,\ldots,\vect{x}_{n})$ and define $p(\vect
{x}_{1:n}|\vectr{\theta})=\prod_{i=1}^n p(\vect{x}_{i}|\vectr{\theta})$.
Therefore the total size of $\vect{x}_{1:n}$ is $nN$. Note that this
could also relate to a stationary time series simulation of length $nN$
(see, e.g., \cite{Gallant2009}). In the case of a stationary time
series or independent and identically distributed (i.i.d.) data it is
not a requirement for the simulated dataset size to be a multiple of
the observed data size. However, for the sake of simplicity we restrict
$n$ to be a positive integer.

Since the likelihood of the auxiliary model is tracta\-ble, we can
potentially consider richly parameterised statistical models to capture
the essential features of the data. We assume throughout the paper that
the dimensionality of the auxiliary model parameter is at least as
large as the dimensionality of the generative model parameter, that is,
$\operatorname{dim}(\vectr{\phi}) \geq\operatorname{dim}(\vectr{\theta})$.
\citet{Jiang2004} note that it still may be possible to obtain useful
estimates of a subset of the parameters when this assumption does not
hold. We do not consider this here.

\section{Approximate Bayesian Computation with Indirect Inference}
\label{sec:ABC II}

\subsection{Approximate Bayesian Computation}\label{s3.1}

ABC is widely becoming a standard tool for performing (approximate)
Bayesian inference on statistical models with computationally
intractable likelihood evaluations but where simulation is
straightforward. ABC analyses set $n=1$ so that the simulated dataset
is the same size as the observed. However, for the purposes of this
paper, we relax this commonly applied restriction for the moment.

\begin{algorithm*}[b]
\caption{MCMC ABC algorithm of \citet{MarjoramEtAl2003}.}
\label{alg:MCMC}
\begin{algorithmic}[1]
\STATE Set $\vectr{\theta}^0$
\STATE Simulate $\vect{x}^0 \sim p(\cdot|\vectr{\theta}^0)$
\FOR{$i=1$ \TO$T$}
\STATE Draw $\vectr{\theta}^* \sim q(\cdot|\vectr{\theta}^{i-1})$
\STATE Simulate $\vect{x}^* \sim p(\cdot|\vectr{\theta}^*)$
\STATE Compute
$r = \min (1,\frac{p(\vectr{\theta}^*)q(\vectr{\theta}^{i-1}|\vectr{\theta}^*)K_\epsilon(\rho(s(\vect{x}^{*}),s(\vect{y})))}{p(\vectr{\theta}^{i-1})q(\vectr{\theta}^*|\vectr{\theta}^{i-1})K_\epsilon(\rho(s(\vect
{x}^{i-1}),s(\vect{y})))} )$
\IF{$\operatorname{uniform}(0,1) < r$}
\STATE$\vectr{\theta}^i = \vectr{\theta}^*$ and $\vect{x}^i = \vect{x}^*$
\ELSE
\STATE$\vectr{\theta}^i = \vectr{\theta}^{i-1}$ and $\vect{x}^i = \vect{x}^{i-1}$
\ENDIF
\ENDFOR
\end{algorithmic}
\end{algorithm*}

In ABC, a summary statistic is defined by a collection of functions
$s_n \dvtx  \mathsf{Y}^n \rightarrow\mathsf{S}$, for each $n \in\mathbb
{N}$. Henceforth, the subscript $n$ is omitted to ease presentation.
Proposed parameter values that produce a simulated summary statistic,
$s(\vect{x}_{1:n})$, ``close'' to the observed summary statistic, $s(\vect
{y})$, are given more weight. Here, we define ``close'' by the
discrepancy function $\rho(s(\vect{x}_{1:n}),s(\vect{y}))$ and a kernel
weighting function, $K_\epsilon(\rho(s(\vect{x}_{1:n}),s(\vect{y})))$,
where $\epsilon$ is a bandwidth referred to as the ABC tolerance. The
ABC target distribution is given by
%
\begin{equation}
\label{eq:ABCTarget} p_{\epsilon,n} (\vectr{\theta}|\vect{y}) \propto p(\vectr{
\theta}) p_{\epsilon,n} (\vect{y}|\vectr{\theta}),
\end{equation}
where
\[
p_{\epsilon,n} (\vect{y}|\vectr{\theta}) \propto \int_{\mathsf
{Y}^n}p(
\vect{x}_{1:n}|\vectr{\theta})K_\epsilon\bigl(\rho\bigl(s(\vect
{x}_{1:n}),s(\vect{y})\bigr)\bigr) \,d\vect{x}_{1:n},
\]
is referred to here as the ABC likelihood. It can be shown that if
$s(\cdot)$ is a sufficient statistic and $n=1$ then $p_{\epsilon,1}
(\vectr{\theta}|\vect{y}) \rightarrow p(\vectr{\theta}|\vect{y})$ as
$\epsilon\rightarrow0$ \citep{Blum2009}. Unfortunately, ABC cannot be
trusted when the value of $n$ is increased in the sense that the target
$p_{\epsilon,n}(\vectr{\theta}|\vect{y})$ can move further away from
$p(\vectr{\theta}|\vect{y})$. In the simple example in Appendix A of the
supplemental article \citep{DrovandiSupp2014}, the posterior
distribution for a univariate $\theta$ converges to a point mass
centred on the single observation $y$ as $n \rightarrow\infty$ and
$\epsilon\rightarrow0$ (see also \cite{Drovandi2012}, pages 28--29, for
a similar example). We also verify this behaviour empirically on a toy
example. This suggests that whilst it is tempting to increase the
simulated dataset size to reduce the variability of the simulated
summary statistic, such an approach is fraught with danger.

Standard ABC procedures can make use of $n$ simulated datasets but in a
different way. Here the ABC likelihood is estimated via $n^{-1} \sum_{i=1}^n K_\epsilon(\rho(s(\vect{x}_{i}),\break s(\vect{y})))$. However, we
note that this is an unbiased estimate (regardless of $n$) of a more
standard ABC likelihood, and thus this process does not alter the ABC
approximation \citep{Andrieu2009}. Therefore, for the remaining
presentation in this section we set $n=1$. However, increasing $n$ may
improve the performance of ABC algorithms and allow a smaller value of
$\epsilon$ for the same computational cost. We do not investigate this here.

How the parameter value is proposed depends on the chosen ABC
algorithm. Here, we use Markov chain Monte Carlo (MCMC ABC, \cite
{MarjoramEtAl2003}) with the proposal distribution $q$ chosen
carefully, and sometimes based on previous ABC analyses. The approach
is shown in Algorithm~\ref{alg:MCMC} for completeness. In the
algorithm, $T$ is the number of iterations. We choose $\vectr{\theta}^0$, the initial value of the chain, so that it is well supported by
the target distribution. This initial value may come from a previous
analysis or by performing some preliminary runs. This algorithm is
mainly used for simplicity, although it is important to note that other
ABC algorithms could be applied.

One difficult aspect in ABC is that often some form of data reduction
is required, to avoid the curse of dimensionality associated with
comparing all the data at once \citep{Blum2009}. This choice of summary
statistic is therefore crucial for a good approximation, even when
$\epsilon$ can be reduced to practically 0. In some applications, it is
possible to propose another model that provides a good description of
the data. The summary statistic used in ABC can be formulated based on
this auxiliary model. These approaches are summarised below.

\subsection{ABC IP}

\citet{DrovandiEtAl2011} suggest using the parameter estimate of the
auxiliary model as the summary statistic to use in ABC. Data are
simulated from the generative model based on a proposed parameter
value, then the auxiliary model parameter is estimated based on this
simulated data. The way the auxiliary parameter is estimated provides a
mapping between the generative model and the auxiliary model
parameters. The ABC algorithm uses a noisy mapping between
$\vectr{\theta}$ and $\vectr{\phi}$ through the simulated data, $\vect{x}$, generated
on the basis of $\vectr{\theta}$, $\vectr{\phi}(\vectr{\theta},\vect{x})$.
For this purpose (explained below), we use the maximum likelihood
estimate (MLE) of the auxiliary model
\[
\vectr{\phi}(\vectr{\theta},\vect{x}) = \arg\max_{\vectr{\phi} \in\Phi
}p_A(
\vect{x}|\vectr{\phi}),
\]
where $\vect{x} \sim p(\cdot|\vectr{\theta})$. ABC IP relies on the
following assumption:

\begin{ass}[(ABC IP Assumptions)]\label{ass:ABCIP}
The estimator of the auxiliary parameter,
$\vectr{\phi}(\vectr{\theta},\vect{x})$, is
unique for all $\vectr{\theta}$ with positive prior support.
\end{ass}

It is important to note that ABC IP (as well as other ABC approaches
below) use $n=1$ so the approximation quality of the method can depend
on the statistical efficiency of the estimator
$\vectr{\phi}(\vectr{\theta},\vect{x})$ based on this finite sample.
Additionally, the MLE is
asymptotically sufficient (\cite{Cox1979}, page~307). For this reason, we
advocate the use of the MLE in general as it is typically more
efficient than other estimators like sample moments (for the auxiliary
model). Sample moments may be computationally easier to obtain, but are
likely to result in a poorer ABC approximation if the statistical
efficiency is lower than the MLE. We note that the optimal choice of
auxiliary estimator (trading off between computational effort and
statistical efficiency) may be problem dependent. An additional
complication is that the auxiliary model is fitted based on data
generated from a different model, $\vect{x} \sim p(\cdot|\vectr{\theta})$.
Therefore, the efficiency of $\vectr{\phi}(\vect{x},\vectr{\theta})$
should be based on $p(\vect{x}|\vectr{\theta})$ not
$p_A(\vect{x}|\vectr{\phi})$ (see, e.g., \cite{Cox1961}).
This can be investigated by simulation.

In Section~\ref{subsec:quantile_example}, we provide an example where
the auxiliary model does not satisfy Assumption~\ref{ass:ABCIP},
creating difficulties for ABC IP. The ABC II methods (and the pdBIL
method) that follow do not necessarily require unique auxiliary
parameter estimates.

An advantage of the II approach to obtaining summary statistics is that
the summary statistic will likely be useful if the auxiliary model fits
the observed data (see Section~\ref{subsec:ABCIISummary} for more
discussion). In the case of ABC IP, the (approximate) covariance matrix
of the auxiliary parameter estimate based on the observed data can be
estimated by the inverse of the observed information matrix [we denote
this information matrix by $\vect{J}(\vectr{\phi}(\vect{y}))$].
Intuitively, we expect this discrepancy function to be more efficient
than a Euclidean distance, as it can take into account the variability
of summary statistics and the correlations between summary statistics.
Denoting the observed summary statistic as $\vectr{\phi}(\vect{y})$ and
the simulated summary statistic as $\vectr{\phi}(\vect{x})$ (dropping
$\vectr{\theta}$ for notational convenience), we use the following
discrepancy for ABC IP:
\begin{eqnarray*}
&&\rho\bigl(s(\vect{x}),s(\vect{y})\bigr)
\\
&&\quad= \sqrt{\bigl(\vectr{\phi}(\vect{x}) -
\vectr{\phi}(\vect{y})\bigr)^T\vect{J}\bigl(\vectr{\phi}(\vect{y})
\bigr) \bigl(\vectr{\phi}(\vect {x}) - \vectr{\phi}(\vect{y})\bigr)}.
\end{eqnarray*}
It is important to note that this is essentially an ABC version of the
classical approach in \citet{Gourieroux1993}. A more appropriate
weighting matrix may involve considering the variance of $\vectr{\phi}(\vect{x},\vectr{\theta})$ when the data are generated under an
alternative model, $\vect{x} \sim p(\cdot|\vectr{\theta})$
(\cite{Cox1961} provides a result, the so-called sandwich estimator).

\subsection{ABC IL}

\citet{Gleim} also describe an approach that uses the auxiliary
likelihood to set up an ABC discrepancy. Here, the ABC discrepancy is
\[
\rho\bigl(s(\vect{x}),s(\vect{y})\bigr) = \log p_A\bigl(\vect{y}|
\vectr{\phi}(\vect {y})\bigr) - \log p_A\bigl(\vect{y}|\vectr{\phi}(
\vect{x})\bigr).
\]
This is effectively an ABC version of the classical approach of \citet
{Smith1993}. We note that $p_A(\vect{y}|\vectr{\phi}(\vect{y}))$ will
remain unchanged throughout the algorithm and provides an upperbound
for values of $p_A(\vect{y}|\vectr{\phi}(\vect{x}))$ obtained for every
simulated dataset. ABC IL uses the same summary statistic as ABC IP but
uses a discrepancy based on the likelihood rather than the Mahalanobis
distance. Note that the discrepancy function for ABC IP appears in the
second-order Taylor series approximation of $\log p_A(\vect{y}|\vectr{\phi}(\vect{x}))$
about $\log p_A(\vect{y}|\vectr{\phi}(\vect{y}))$
(\cite{Davison2003}, page~126) assuming standard regularity conditions
for $p_A(\vect{y}|\vectr{\phi}(\vect{y}))$ and $\vectr{\phi}(\vect{y})$.
The ABC tolerance could be viewed here as a certain cut-off value of
the auxiliary log-likelihood. The ABC IL approach relies on the
following assumption.

\begin{ass}[(ABC IL Assumptions)]\label{ass:ABCIP_likelihood}
The auxiliary likelihood evaluated at the auxiliary estimate, $p_A(\vect
{y}|\vectr{\phi}(\vect{x},\vectr{\theta}))$, is unique for all
$\vectr{\theta}$ with positive prior support.
\end{ass}

We note that this assumption can still be satisfied even when the
auxiliary model does not have a unique MLE (see Section~\ref{subsec:quantile_example} for an example).

The ABC IP and ABC IL methods use parameter estimates of the auxiliary
model as summary statistics and can thus be expensive as it can involve
a numerical optimisation every time data is simulated from the
generative model. The next approach to obtaining summary statistics
from II avoids this optimisation step.

\subsection{ABC IS}

\citet{Gleim} advocate the use of the score vector of the auxiliary
model evaluated at the auxiliary
MLE, $\vectr{\phi}(\vect{y})$, as the summary statistic. We denote the
score vector of the auxiliary model as
\[
\vect{S}_A(\vect{y},\vectr{\phi}) = \biggl(\frac{\partial\log p_A(\vect
{y}|\vectr{\phi})}{\partial\phi_1},
\ldots,\frac{\partial\log
p_A(\vect{y}|\vectr{\phi})}{\partial\phi_{\operatorname{dim}(\vectr{\phi})}} \biggr)^T,
\]
where $\vectr{\phi} = (\phi_{1},\ldots,\phi_{\operatorname{dim}(\vectr{\phi})})^T$.

Each component of the summary statistic involving the observed data and
the MLE, $\vect{S}_A(\vect{y},\vectr{\phi}(\vect{y}))$, is assumed to be
numerically 0 under standard regularity assumptions (see below,
Assumption~\ref{ass:ABCIS}). Thus, the search is for parameter values
of the generative model that lead to simulated data, $\vect{x}$, that
produces a score close to $\vect{0}$. Noting that the approximate
covariance matrix of the observed score is given by the observed
information matrix $\vect{J}(\vectr{\phi}(\vect{y}))$, the following ABC
discrepancy is obtained for ABC IS
\begin{eqnarray*}
&& \rho\bigl(s(\vect{x}),s(\vect{y})\bigr)
\\
&&\quad = \sqrt{\vect{S}_A\bigl(
\vect{x},\vectr{\phi}(\vect{y})\bigr)^T \vect{J}\bigl(\vectr{\phi}(
\vect{y})\bigr)^{-1}\vect{S}_A\bigl(\vect {x},\vectr{\phi}(
\vect{y})\bigr)}.
\end{eqnarray*}
This is essentially an ABC version of \citet{Gallant1996}.

This approach is fast relative to ABC IP when the MLE of the auxiliary
model is not analytic whilst the score is analytic since no numerical
optimisation is required every time data are simulated from the
generative model. Of course, it may be necessary to estimate the score
numerically, which would add another layer of approximation and may be
slower. In the examples of this paper, we are able to obtain the score
analytically. ABC IS relies on the following assumptions.

\begin{ass}[(ABC IS Assumptions)]\label{ass:ABCIS}
The MLE of the auxiliary model fitted to the observed data,
$\vectr{\phi}(\vect{y})$, is an interior point of the
parameter space of $\vectr{\phi}$ and $\vect{J}(\vectr{\phi}(\vect{y}))$
is positive definite. The
log-likelihood of the auxiliary model,
$\log p_A(\vectr{\cdot}|\vectr{\phi})$, is differentiable and the score,
$\vect{S}_A(\vect{x},\vectr{\phi}(\vect{y}))$, is unique for
any $\vect{x}$ that may be drawn according
to any $\vectr{\theta}$ that has positive prior support.
\end{ass}

We note that Assumption~\ref{ass:ABCIS} is generally weaker than
Assumption~\ref{ass:ABCIP} (ABC IP), since it may still hold even if
the MLE of the auxiliary model is not unique.


\subsection{Discussion on ABC II Summary Statistics} \label{subsec:ABCIISummary}

Only models in the exponential family possess a minimal sufficient
statistic with dimension equal to that of $\operatorname{dim}(\vectr{\theta})$. For other models, under suitable conditions, the
Pitman--Koopman--Darmois theorem states that the dimension of any
sufficient statistic increases with the sample size. For many complex
models, such as those considered in the ABC setting, the minimal
sufficient statistic will be the full dataset (or the full set of order
statistics if the data are i.i.d.). The summary statistic produced by
ABC II will always have dimension $\operatorname{dim}(\vectr{\phi})$, and thus
will not produce sufficient statistics in general (this argument of
course carries over to any ABC method that uses some data reduction
technique; see \cite{Blum2013}, for a review). Intuitively, our
suggestion is that the summary statistic produced by ABC II should
carry most of the information contained in the observed data provided
that the auxiliary model provides a good description of the data.
Unfortunately, this is difficult to verify since it is usually not
possible to quantify the amount of information lost in data reduction.
Despite this, by conducting goodness-of-fit tests and/or residual
analysis on the auxiliary model fit to the data will at least provide
some guidance on the usefulness of the summary statistic produced by
ABC II. This is in contrast to the more traditional approach of
summarising based on simple functions of the data (e.g., \cite
{DrovandiPettitt2011}), whose utility is difficult to assess prior to
running an ABC analysis without performing an expensive simulation
study. Furthermore, ABC II methods provide natural discrepancy
functions between summary statistics as shown above. Selecting the
discrepancy function and determining appropriate weighting of the
summary statistics in traditional ABC can be problematic.

It is well known that the choice of summary statistic in ABC involves a
compromise between \mbox{sufficiency} and dimensionality \citep{Blum2013}. A
low-dimensional and near-sufficient summary statistic represents an
optimal trade-off. Another advantage of ABC II over usual ABC is the
dimensionality of the ABC II summary statistic can be controlled by
selecting parsimonious auxiliary models and using standard model choice
techniques to choose between a set of possible auxiliary models [e.g.,
the Akaike information criterion (AIC) and the Bayesian information
criterion (BIC)].

\section{Parametric Bayesian Indirect Likelihood (\lowercase{p}BIL)}\label{sec:BIL}

\subsection{Parametric Bayesian Indirect Likelihood for the Full Data
(pdBIL)} \label{subsec:BIL_target}

\citet{Reeves2005} and \citet{Gallant2009} propose a method that has
similar steps to ABC IP and ABC IL but is theoretically quite
different, as we show below. After data are simulated from the
generative model, the auxiliary parameter is estimated. This auxiliary
estimate is then passed into the auxiliary likelihood of the observed
data. This likelihood is then treated in the usual way and fed into a
Bayesian algorithm, for example, MCMC. One first defines a collection
of functions $\vectr{\phi}_n\dvtx \Theta\times\mathsf{Y}^n \rightarrow\Phi
$. The artificial likelihood is then defined as follows:
\[
p_{A,n}(\vect{y}|\vectr{\theta})=\int_{\mathsf{Y}^{n}}p_{A}
\bigl(\vect {y}|\vectr{\phi}_{n}(\vectr{\theta},\vect{x}_{1:n})
\bigr)\prod_{i=1}^{n}p(\vect
{x}_{i}|\vectr{\theta})\,d\vect{x}_{1:n},
\]
and the target distribution of this approach is given by
\[
p_{A,n}(\vectr{\theta}|\vect{y}) \propto p_{A,n}(\vect{y}|
\vectr{\theta})p(\vectr{\theta}),
\]
where the subscripts $A$ and $n$ denote the dependence of the target on
the auxiliary model choice and the number of replicate simulated
datasets, respectively. This approach is effectively a Bayesian version
of the simulated quasi-maximum likelihood approach of \citet
{Smith1993}. \citet{Smith1993} proposes to maximise $p_{A,n}(\vect
{y}|\vectr{\theta})$ with respect to $\vectr{\theta}$. Instead of
applying this as an ABC discrepancy as in the ABC IL method above,
\citet{Reeves2005} and \citet{Gallant2009} treat this auxiliary
likelihood as a replacement to the likelihood of the generative model
in the same way that \citet{Smith1993} does.

It is important to note that this approach does not perform a
comparison of summary statistics, and hence there is no need to choose
an ABC tolerance. Thus, it is not a standard ABC algorithm. We refer to
this approach simply as pdBIL, since we apply a ``parametric'' auxiliary
model for the full ``data''. When the full data have been summarised as a
summary statistic, $s(\vect{y})$, an alternative approach is to apply a
parametric auxiliary model for the summary statistic likelihood,
$p(s(\vect{y})|\vectr{\theta})$ (see Section~\ref{subsec:psBIL} for more
details). In Figure~\ref{fig:BIL}, these methods fall under the pBIL
class. Within this class, the focus of this paper is on the pdBIL method.

\subsubsection{The pdBIL approximation}

The theoretical aspects of this approach are yet to be investigated in
the literature. Some clues are offered in \citet{Reeves2005} and \citet
{Gallant2009}, but we formalise and extend the theory here. The
subscript $n$ is used to denote that this target remains an
approximation to the true posterior distribution and that the
approximate target may change with $n$ (we show below that in general
the target does depend on $n$). However, because it is not an ABC
algorithm, it is unclear how pdBIL behaves as $n$ increases. \citet
{Gallant2009} use a very large simulation size ($n \approx700$),
without a theoretical investigation.

With $\vect{y}$ fixed, we consider a potential limiting likelihood
$p_A(\vect{y}|\vectr{\phi}(\vectr{\theta}))$ with associated posterior
\[
p_{A}(\vectr{\theta}|\vect{y})\propto p_{A}\bigl(\vect{y}|
\vectr{\phi}(\vectr{\theta})\bigr)p(\vectr{\theta}).
\]
Note that the parameter of $p_{A,n}(\vect{y}|\vectr{\theta})$ is $\vectr{\theta}\in\Theta$ but the parameter of $p_A(\vect{y}|\vectr{\phi}(\vectr{\theta}))$ is $\vectr{\phi}(\vectr{\theta})\in\Phi$. We emphasise that
the results below all assume that $\vect{y}$ is a fixed value in
$\mathsf{Y}$.

To ease presentation, we define the random variable $\vectr{\phi}_{\vectr{\theta},n}=\vectr{\phi}_{n}(\vectr{\theta},\vect{X}_{1:n})$ where $(\vect
{X}_{i})$ is a sequence of i.i.d. random variables distributed
according to $p(\cdot|\vectr{\theta})$,
and we can write
\[
p_{A,n}(\vect{y}|\vectr{\theta})=\mathsf{E} \bigl[p_{A}(
\vect{y}|\vectr{\phi}_{\vectr{\theta},n}) \bigr],
\]
where $\mathsf{E}$ is expectation with respect to the distribution of
$\vect{X}_{1:n}$.

The results below provide sufficient conditions under which, as $n
\rightarrow\infty$, $p_{A,n}(\cdot| \vect{y}) \rightarrow p_{A}(\cdot
|\vect{y})$
pointwise and
\[
\int_{\Theta}f(\vectr{\theta})p_{A,n}(\vectr{\theta} |
\vect{y})\,d\vectr{\theta} \rightarrow\int_{\Theta}f(\vectr{
\theta})p_{A}(\vectr{\theta} | \vect{y})\,d\vectr{\theta},
\]
where $f$ is some function of $\vectr{\theta}$ whose posterior
expectation is of interest. This does not assume that $p_{A}(\cdot|
\vect{y})=p(\cdot| \vect{y})$, which in general will not be the case.
A useful tool to allow us to answer both questions is provided by
\citet{Billingsley2009}.

\begin{them}[(\cite{Billingsley2009}, Theorem~3.5)]\label{thm:billingsley}
If $X_{n}$ is a sequence of uniformly integrable random variables and
$X_{n}$ converges in distribution to $X$ then $X$ is integrable and
$\mathsf{E}X_{n}\rightarrow\mathsf{E}X$.
\end{them}

\begin{rem}
A simple sufficient condition for uniform integrability is that for
some $\delta>0$,
\[
\sup_{n}\mathsf{E} \bigl(|X_{n}|^{1+\delta}
\bigr)<\infty.
\]
\end{rem}

\begin{res}
\label{res:conv_posterior_and_expectation}
Assume that $p_{A,n}(\vect{y}|\vectr{\theta})\rightarrow p_{A}(\vect
{y}|\vectr{\phi}(\vectr{\theta}))$
as $n\rightarrow\infty$ for all $\vectr{\theta}$ with positive prior
support, $\inf_{n}\int_{\Theta}p_{A,n}(\vect{y} | \vectr{\theta})
p(\vectr{\theta})\,d\vectr{\theta}>0$ and $\sup_{\vectr{\phi}\in\Phi}p_{A}(\vect
{y}|\vectr{\phi})<\infty$. Then
\[
\lim_{n\rightarrow\infty}p_{A,n}(\vectr{\theta}|
\vect{y})=p_{A}(\vectr{\theta}|\vect{y}).
\]
Furthermore, if $f\dvtx \Theta\rightarrow\mathbb{R}$ is a continuous function
satisfying $\sup_{n}\int_{\Theta}|f(\vectr{\theta})|^{1+\delta
}p_{A,n}(\vectr{\theta}|\vect{y})\,d\vectr{\theta}<\infty$
for some $\delta>0$ then
\[
\lim_{n\rightarrow\infty}\int_{\Theta}f(\vectr{
\theta})p_{A,n}(\vectr{\theta}|\vect{y})\,d\vectr{\theta}=\int
_{\Theta}f(\vectr{\theta})p_{A}(\vectr{\theta}|
\vect{y})\,d\vectr{\theta}.
\]
\end{res}

\begin{pf}
The first part follows from the fact that the numerator of
\[
p_{A,n}(\vectr{\theta}|\vect{y})=\frac{p_{A,n}(\vect{y}|\vectr{\theta})p(\vectr{\theta})}{\int_{\Theta}p_{A,n}(\vect{y}|\vectr{\theta})p(\vectr{\theta})\,d\vectr{\theta}}
\]
converges pointwise and the denominator is positive and converges
by the bounded convergence theorem. For the second part, if for each
$n\in\mathbb{N}$, $\vectr{\theta}_{n}$ is distributed according to
$p_{A,n}(\cdot|\vect{y})$
and $\vectr{\theta}$ is distributed according to $p_{A}(\cdot|\vect{y})$
then $\vectr{\theta}_{n}$
converges to $\vectr{\theta}$ in distribution as $n\rightarrow\infty$ by
Scheff\'{e}'s lemma \citep{Scheffe1947}. Since $f$ is continuous,
$f(\vectr{\theta}_{n})$ converges in distribution to $f(\vectr{\theta})$
as $n\rightarrow\infty$ by the continuous mapping theorem and we
conclude by application of Theorem~\ref{thm:billingsley}.
\end{pf}

A simple condition for $p_{A,n}(\vect{y}|\vectr{\theta})\rightarrow
p_{A}(\vect{y}|\vectr{\phi}(\vectr{\theta}))$
as $n\rightarrow\infty$ to hold is provided by the following result.

\begin{res}\label{res:conv_likelihood}
Assume that $p_{A}(\vect{y}|\vectr{\phi}_{\vectr{\theta},n})$ converges
in probability to $p_{A}(\vect{y} | \vectr{\phi}(\vectr{\theta}))$ as
$n\rightarrow\infty$. If
\[
\sup_{n}\mathsf{E} \bigl[\bigl |p_{A}(\vect{y}|\vectr{
\phi}_{\vectr{\theta},n})\bigr |^{1+\delta} \bigr]<\infty
\]
for some $\delta>0$ then $p_{A,n}(\vect{y}|\vectr{\theta})\rightarrow
p_{A}(\vect{y}|\vectr{\phi}(\vectr{\theta}))$ as $n\rightarrow\infty
$.
\end{res}

\begin{pf}
The result follows by application of Theorem~\ref{thm:billingsley}.
\end{pf}

Although the results above hold under conditions on the fixed, observed
data $\vect{y}$, they will often hold for a range of possible values of
$\vect{y}$.

The function $\vectr{\phi}(\vectr{\theta})$ is often referred to as the
mapping or binding function in the II literature. In \citet
{Gallant2009}, it is assumed that this function is 1--1 but the results
above demonstrate that this is not a necessary condition for Result~\ref
{res:conv_likelihood} to hold. The following example where the
auxiliary model is a mixture model demonstrates this principle.

Assume that the true model is $N(\vect{y};\theta,1)$ while the
auxiliary model is a mixture of normals, $wN(\vect{y};\theta_1,\allowbreak  1) +
(1-w)N(\vect{y};\theta_2,1)$ so that $\vectr{\phi} = (\theta_1,\theta
_2,w)$. Assuming an infinite sample from the true model, there are an
infinite number of MLEs of the auxiliary model; $\vectr{\phi}(\theta) =
(\theta,\theta,w)$ where $0 \leq w \leq1$, $\vectr{\phi}(\theta) =
(\theta,\theta_2,1)$ where $-\infty< \theta_2 < \infty$ or $\vectr{\phi}(\theta) = (\theta_1,\theta,0)$ where $-\infty< \theta_1 < \infty$.
All of these possible mappings produce the same value of the auxiliary
likelihood, which coincides with the value of the generative likelihood.

It is straightforward under the assumptions of Results~\ref
{res:conv_posterior_and_expectation}--\ref{res:conv_likelihood} to show
that pdBIL will target the true posterior as $n \rightarrow\infty$ if
the true model is contained within the auxiliary model. When the
generative model is a special case of the auxiliary model, using the
notation of \citet{Cox1990}, the auxiliary and generative parameter can
be written as $\vectr{\phi} = (\vectr{\theta}_{e},\vectr{\gamma})$
and $\vectr{\theta} = (\vectr{\theta}_{r},\vectr{\gamma}_{\vectr{0}})$ where $e$ and $r$ denote
``extended'' and ``reduced'' respectively and $\vectr{\gamma}_{\vectr{0}}$ is fixed.
The proof of this result is straightforward. It involves demonstrating
that $\vectr{\phi}(\vectr{\theta},\vect{x}_{1:n})$ is consistent also for
the parameter of the generative (reduced) model. Therefore, when $n
\rightarrow\infty$ the generative and auxiliary likelihoods will coincide.

This theoretical result cannot typically be realised in practice since
a model which incorporates an intractable model as a special case is
likely to also be intractable. However, it does suggest that the
auxiliary model be chosen to be adequately flexible to give a good
approximation of the generative likelihood for $\vectr{\theta}$ values
with positive prior support. In practice, our empirical evidence
indicates that it is only necessary for the auxiliary likelihood to
mimic the behaviour of the generative likelihood for the values of
$\vectr{\theta}$ with nonnegligible posterior support and for the
auxiliary likelihood to be negligible in regions of the parameter space
with little posterior support. If this is not the case, it is likely
that the pdBIL method will lead to poor approximations (as we
demonstrate in Section~\ref{subsec:toy_example}).

If the binding function, $\vectr{\phi}(\vectr{\theta})$, were known, the
pdBIL method would proceed straightforwardly. Since it will not be
available in practice, it can be estimated indirectly through the
simulated data $\vect{x}_{1:n}$. From the above, it is desirable for
the pdBIL method if the auxiliary likelihood is as close as possible to
the true likelihood. \citet{Gallant2009} show, for a particular choice
of the auxiliary model, that choosing the MLE for $\vectr{\phi}(\vectr{\theta},\vect{x}_{1:n})$ minimises the Kullback--Leibler divergence
between the generative and auxiliary likelihoods. Furthermore, this
choice will often lead to Results~\ref
{res:conv_posterior_and_expectation}--\ref{res:conv_likelihood}
holding. Therefore, we advocate the use of the MLE with this method.
When the MLE of the auxiliary model is used, \citet{Cox1961} provides
an expression which defines $\vectr{\phi}(\vectr{\theta})$ [see equation
(25) of \cite{Cox1961}].

\begin{algorithm*}[b]
\caption{MCMC pdBIL algorithm (see also \cite{Gallant2009}).}
\label{alg:MCMC-BIL}
\begin{algorithmic}[1]
\STATE Set $\vectr{\theta}^0$
\STATE Simulate $\vect{x}_{1:n}^* \sim p(\cdot|\vectr{\theta}^0)$
\STATE Compute $\vectr{\phi}^0 = \arg\max_{\vectr{\phi} \in\Phi
}p_A(\vect{x}_{1:n}^*|\vectr{\phi})$

\FOR{$i=1$ \TO$T$}
\STATE Draw $\vectr{\theta}^* \sim q(\cdot|\vectr{\theta}^{i-1})$
\STATE Simulate $\vect{x}_{1:n}^* \sim p(\cdot|\vectr{\theta}^*)$
\STATE Compute $\vectr{\phi}(\vect{x}_{1:n}^*) = \arg\max_{\vectr{\phi}}p_A(\vect{x}_{1:n}^*|\vectr{\phi})$
\STATE Compute
$r = \min (1,\frac{p_A(\vect{y}|\vectr{\phi}(\vect
{x}_{1:n}^*))p(\vectr{\theta}^*)q(\vectr{\theta}^{i-1}|\vectr{\theta}^*)}{p_A(\vect{y}|\vectr{\phi}^{i-1})p(\vectr{\theta}^{i-1})q(\vectr{\theta}^*|\vectr{\theta}^{i-1})} )$
\IF{$\operatorname{uniform}(0,1) < r$}
\STATE$\vectr{\theta}^i = \vectr{\theta}^*$
\STATE$\vectr{\phi}^{i} = \vectr{\phi}(\vect{x}_{1:n}^*)$
\ELSE
\STATE$\vectr{\theta}^i = \vectr{\theta}^{i-1}$
\STATE$\vectr{\phi}^{i} = \vectr{\phi}^{i-1}$
\ENDIF
\ENDFOR
\end{algorithmic}
\end{algorithm*}

It remains to be seen what the target of $p_{A,n}(\vectr{\theta}|\vect
{y})$ is relative to $p_{A}(\vectr{\theta}|\vect{y})$. From an intuitive
perspective, increasing $n$ leads to a more precise determination of
the mapping $\vectr{\phi}(\vectr{\theta})$, and thus should lead to a
better approximation. A more theoretical argument is as follows. The
approximations will\vadjust{\goodbreak} coincide with each other provided $\mathsf
{E}[p_A(\vect{y}|\vectr{\phi}_{\vectr{\theta},n})] =
p_A(\vect{y}|\vectr{\phi}(\vectr{\theta}))$ \citep{Andrieu2009}. Unfortunately, even if one
uses an unbiased estimator for the auxiliary parameter, that is $\mathsf
{E}[\vectr{\phi}_{\vectr{\theta},n}] = \vectr{\phi}(\vectr{\theta})$ for
any value of $n$, this result will still rarely hold in general. The
likelihood can be viewed here as a nonlinear function of the auxiliary
parameter estimate, and so
$\mathsf{E}[p_A(\vect{y}|\vectr{\phi}_{\vectr{\theta},n})] \neq p_A(\vect
{y}|\vectr{\phi}(\vectr{\theta}))$ in general. Our empirical evidence
(see Section~\ref{sec:Results}) suggests that $p_{A,n}(\vectr{\theta}|\vect{y})$
typically becomes less precise relative to $p_{A}(\vectr{\theta}|\vect{y})$
as the likelihood estimate becomes more noisy, that
is, when $n$ is reduced. The message of Results~\ref
{res:conv_posterior_and_expectation}--\ref{res:conv_likelihood} is
that, provided the auxiliary model is suitably chosen, a better
approximation can be anticipated by taking $n$ as large as possible,
which has the effect of reducing the bias of $p_A(\vect{y}|\vectr{\phi}_{\vectr{\theta},n})$.

\subsubsection{MCMC pdBIL} \label{subsec:MCMC_BIL}

As an example, MCMC can be used to sample from the pdBIL target
(referred to here as MCMC pdBIL, see \cite*{Gallant2009}). This
approach is presented in Algorithm~\ref{alg:MCMC-BIL}.

As Results~\ref{res:conv_posterior_and_expectation}--\ref
{res:conv_likelihood} suggest, the aim with pdBIL is to select $n$ as
large as possible. We demonstrate in the examples that it is desirable
to consider values of $n$ greater than one due to the improved
statistical efficiency of MCMC pdBIL (and potentially other algorithms
that implement pdBIL) when increasing $n$. Of course, the method will
become computationally infeasible for very large $n$.

\subsection{Parametric Bayesian Indirect Likelihood for a Summary
Statistic (psBIL)} \label{subsec:psBIL}

\citet{Wood2010} proposes a method called synthetic likelihood when it
is convenient to perform inference on the basis of a set of summary
statistics rather than the full data. Considering Bayesian inference,
the target distribution when the data have been reduced to a summary
statistic is given by
\[
p\bigl(\vectr{\theta}|s(\vect{y})\bigr) \propto p\bigl(s(\vect{y})|\vectr{
\theta}\bigr)p(\vectr{\theta}).
\]
The major issue with this construction is that there is no analytical
form for the likelihood function of the summary statistic, $p(s(\vect
{y})|\vectr{\theta})$. \citet{Wood2010} overcomes this by applying,
based on the terminology in this paper, a parametric auxiliary model
for this probability distribution, $p_A(s(\vect{y})|\vectr{\phi}(\vectr{\theta}))$. In our framework, an approach that applies a parametric
auxiliary model to form the likelihood of the summary statistic rather
than the likelihood of the full data (as is presented in Section~\ref{subsec:BIL_target}) is referred to here as psBIL (where ``s'' denotes
``summary statistic''). Therefore the Bayesian version of \citet
{Wood2010} is a psBIL approach. For the auxiliary likelihood,
$p_A(s(\vect{y})|\vectr{\phi}(\vectr{\theta}))$, \citet{Wood2010}
considers using the likelihood of a multivariate normal distribution,
$N(\vectr{\mu}(\vectr{\theta}),\vectr{\Sigma}(\vectr{\theta}))$, where
$\vectr{\phi}(\vectr{\theta}) = (\vectr{\mu}(\vectr{\theta}),\vectr{\Sigma}(\vectr{\theta}))$ is the auxiliary parameter. As is the case with
pdBIL, the binding function $\vectr{\phi}(\vectr{\theta})$ is rarely
known but can be estimated via simulated data from the generative model
for a particular value of $\vectr{\theta}$. Using our notation, we
obtain the following dataset from the true model of the summary
statistic, $(s(\vect{x}_{1}),\ldots,s(\vect{x}_{n}))$. This represents
$n$ i.i.d. observations from $s(\cdot)|\vectr{\theta}$. An advantage of
selecting such a simple auxiliary model is that the MLE has the
analytic form
\begin{eqnarray*}
\vectr{\mu}(\vect{x}_{1:n},\vectr{\theta}) &=& \frac{1}{n}\sum
_{i=1}^n s(\vect{x}_{i}),
\\
\vectr{\Sigma}(\vect{x}_{1:n},\vectr{\theta}) &=& \frac{1}{n}
\sum_{i=1}^n \bigl(s(\vect{x}_{i})-
\vectr{\mu}(\vect{x}_{1:n},\vectr{\theta})\bigr)
\\
&&\phantom{\frac{1}{n}
\sum_{i=1}^n \bigl(}{}\cdot\bigl(s(\vect
{x}_{i})-\vectr{\mu}(\vect{x}_{1:n},\vectr{\theta})
\bigr)^T,
\end{eqnarray*}
where the superscript $T$ denotes transpose. The auxiliary likelihood
used is then based on $N(\vectr{\mu}(\vect{x}_{1:n},\vectr{\theta}),\allowbreak
\vectr{\Sigma}(\vect{x}_{1:n},\vectr{\theta}))$. Our results indicate that the
target distribution of this method will depend on $n$, and, if the
auxiliary model for the summary statistic likelihood is reasonable,
better approximations of $p(\vectr{\theta}|s(\vect{y}))$ are likely to
be obtained for large $n$.

\section{ABC as a BIL Method with Nonparametric Auxiliary Model} \label{sec:ABCasBIL}

An alternative and perhaps natural candidate for $p_A$ is to use a
kernel density estimate based on the samples $\vect{x}_{1:n}$. This
corresponds to choosing $\vectr{\phi}(\vectr{\theta},\vect{x}_{1:n})=\vect
{x}_{1:n}$ and we define
\[
p_A\bigl(\vect{y}|\vectr{\phi}(\vectr{\theta},\vect{x}_{1:n})
\bigr) = p_A(\vect {y}|\vect{x}_{1:n}) = \frac{1}{n}
\sum_{i=1}^{n} K_\epsilon\bigl(\rho(
\vect {y},\vect{x}_{i})\bigr),
\]
\citep{Diggle1984}, where $\epsilon$ is the bandwidth parameter. We have then
\begin{eqnarray*}
p_{A,n}(\vect{y}|\vectr{\theta}) &=& \int_{\mathsf{Y}^n}p_A
\bigl(\vect{y}|\vectr{\phi}(\vectr{\theta},\vect{x}_{1:n})\bigr) \prod
_{j=1}^n p(\vect{x}_j|\vectr{
\theta})\,d\vect{x}_{1:n}
\\
&=& \int_{\mathsf{Y}^n}\frac{1}{n}\sum
_{i=1}^{n} K_\epsilon\bigl(\rho(\vect {y},
\vect{x}_{i})\bigr)\prod_{j=1}^n
p(\vect{x}_j|\vectr{\theta})\,d\vect {x}_{1:n}
\\
&=& \frac{1}{n}\sum_{i=1}^{n}\int
_{\mathsf{Y}^n} K_\epsilon\bigl(\rho(\vect {y},
\vect{x}_{i})\bigr)\prod_{j=1}^n
p(\vect{x}_j|\vectr{\theta})\,d\vect {x}_{1:n}
\\
&=& \int_{\mathsf{Y}} K_\epsilon\bigl(\rho(\vect{y},\vect{x})
\bigr)p(\vect{x}|\vectr{\theta})\,d\vect{x}
\\
&\equiv& p_\epsilon(\vect{y}|\vectr{\theta}),
\end{eqnarray*}
and this is exactly the form of the standard ABC likelihood. In
addition, $n$ does not affect the likelihood (although it may help
computationally in some algorithms) and $\epsilon$ controls the level
of approximation. Here, we see that this is an estimate of the ABC
likelihood where the comparison is made between the full datasets.
Here, we obtain the npdBIL approach as presented in Figure~\ref{fig:BIL}
(where ``d'' corresponds to full ``data''). Alternatively, a
nonparametric density estimate of the auxiliary model of the summary
statistic likelihood, $p_A(\vect{s}(\vect{y})|\vectr{\phi}(\vect
{x}_{1:n},\vectr{\theta}))$, could be applied. Using a similar procedure
to above, we obtain $p_\epsilon(s(\vect{y})|\vectr{\theta})$. We refer
to this in Figure~\ref{fig:BIL} as npsBIL (npBIL based on a ``summary
statistic''). This is the approach adopted by \citet{Creel2013},
however, their focus is on point estimation (posterior mean).
Unfortunately, \citet{Creel2013} refer to their Bayesian estimator as
BIL, however, under our framework BIL is a more general class of
methods. Of course, if the summary statistic is derived from some
parametric auxiliary model, then the ABC II class of method is
recovered as an npsBIL method. The reader is again referred to
Figure~\ref{fig:BIL} to see the connection between these methods.

By selecting a parametric model for the auxiliary likelihood (pBIL), we
can potentially overcome the curse of dimensionality associated with
the nonparametric aspect of ABC. This requires further research. Of
course, finding a suitable parametric auxiliary model may be
challenging in practice.

\section{Comparison of ABC II and \lowercase{pd}BIL} \label{sec:compareBII}

There are a few remarks to be made about the above results in relation
to theoretical comparisons between ABC II and pdBIL.

\begin{rem}
\label{rem:BIL_increasing_n}
Under suitable conditions better approximations with pdBIL are obtained
by increasing $n$. This is in stark contrast with ABC II, which cannot
be trusted for $n>1$.
\end{rem}

\begin{rem}
\label{rem:ABC_never_correct}
In the case where the true model is a special case of the auxiliary
model, the pdBIL method will be exact in the limit as $n \rightarrow
\infty$. In contrast, in this ideal situation, ABC II still does not
produce sufficient statistics (see the dimensionality argument in
Section~\ref{subsec:ABCIISummary}) and will not target the true
posterior in the limit as $\epsilon\rightarrow0$. An example is where
the true model is a t-distribution with location, scale and degrees of
freedom of $\mu$, $\sigma$ and 1, respectively. The auxiliary model is
a more general t-distribution with degrees of freedom~$\nu$. In this
case, the pdBIL method is exact in the limit as $n \rightarrow\infty$
as the true model is incorporated within the auxiliary model.
Unfortunately, ABC II does not produce a sufficient statistic as the
summary statistic will be of dimension three whilst it is known for
this model the minimal sufficient statistic consists of all the order
statistics. Of course, finding an auxiliary model that satisfies this
condition in practice will rarely be feasible.
\end{rem}

\begin{rem}
\label{rem:BIL_never_correct}
Even if the auxiliary parameter estimate or score happen to be a
sufficient statistic for the generative model, pdBIL still will not
generally target the true posterior, as the auxiliary and generative
likelihoods will still not match up. In this situation, the ABC II
approaches will enjoy convergence to the true posterior as $\epsilon
\rightarrow0$ whilst pdBIL will not converge to the true posterior as
$n \rightarrow\infty$. However, sufficient statistics are rarely
achieved in practice.
\end{rem}

Remarks~\ref{rem:ABC_never_correct} and~\ref{rem:BIL_never_correct}
demonstrate that pBII methods generally will not (and rarely will)
target the true posterior distribution asymptotically. This is
generally the case for other techniques in the literature for dealing
with models that have intractable likelihood functions. There are some
exceptions to this. For example, exact techniques are available for
so-called doubly intractable models when perfect simulation from the
generative model is possible (e.g., \cite{Moller2006}; \cite{Murray2006}).
Furthermore, so-called pseudo-marginal methods \citep{Andrieu2009} are
applicable when a positive and unbiased estimator of the likelihood is
available and is a current area of research. Despite not being exact,
we demonstrate that pBII methods can produce quite good approximations
in some applications.

The characteristics of a good auxiliary model differ between the ABC II
and pdBIL methods. In the context of ABC II, we simply require a good
summarisation of the data, that is, a low-dimensional summary statistic
that hopefully carries most of the information in the observed data.
Therefore, we feel that it is useful if the auxiliary model in this
context provides a good fit to the data and is parsimonious, so that
the essential features of the data are described well and as succinctly
as possible. This is independent of the process for selecting a
generative model. Therefore, the same auxiliary model should be used
regardless of which generative model is fitted to the data. For pdBIL,
we require a flexible auxiliary model that can mimic the behaviour of
the generative model for different values of $\vectr{\theta}$ within the
posterior support. Here, it is not necessary for the auxiliary model to
provide a good fit to the data considering the fact that the generative
model being proposed might be mis-specified. The auxiliary model chosen
for pdBIL may alter depending on the generative model being proposed.
In our examples, the generative model is either known or provides a
good fit to the data. In such cases, it would not be uncommon to choose
the same auxiliary model for the ABC II and pdBIL methods.

The conditions required for pdBIL to produce exact results are very
strong and finding an auxiliary model that is sufficiently flexible so
that the auxiliary likelihood can mimic the generative likelihood could
be difficult in practice. In some applications, an auxiliary model that
is a simplified version of the generative model may be specified where
the parameter of each model has the same interpretation. For example,
the auxiliary model for a continuous time Markov jump process may be
its corresponding linear noise approximation. In such situations, the
pdBIL method is unlikely to perform well whilst it remains possible
that such an approximate model could produce \mbox{useful} summary statistics
for ABC even though the auxiliary model would not fit the data well.
\citet{Jiang2004} show that II can work well in the classical framework
when the auxiliary model is a simplified version of the generative
model. Further research is required in the Bayesian setting.

An additional advantage of the ABC II approach over pdBIL is the extra
flexibility of being able to accommodate additional summary statistics
that do not involve an auxiliary model, since this method belongs in
the more general npsBIL class (see \cite{Wood2010}, for an example
where the summary statistic is a combination of auxiliary parameter
estimates and other summary statistics). \citet{Jiang2004} and \citet
{Heggland2004} consider II applications in a classical framework where
the comparison of observed and simulated data is made on the basis of
both an auxiliary model and supplementary summary statistics.

\section{Examples} \label{sec:Results}

\subsection{Toy Example} \label{subsec:toy_example}

In this example, we consider a simple model so that exact Bayesian
inferences are trivially obtained. Our intention here is to investigate
the theoretical considerations in Section~\ref{sec:BIL}. In particular,
we show that when the auxiliary model is reasonable, pdBIL produces
better approximations as the size of simulated datasets goes beyond
that of the observed data and as a useful by-product increases the
acceptance probability of the MCMC moves. We also demonstrate
empirically that unfortunately ABC approaches (including those using II
to obtain summary statistics) do not possess this same desirable
property as $n$ is increased. Additionally, we investigate the output
of pdBIL when the auxiliary model is poorly chosen.

\begin{figure*}

\includegraphics{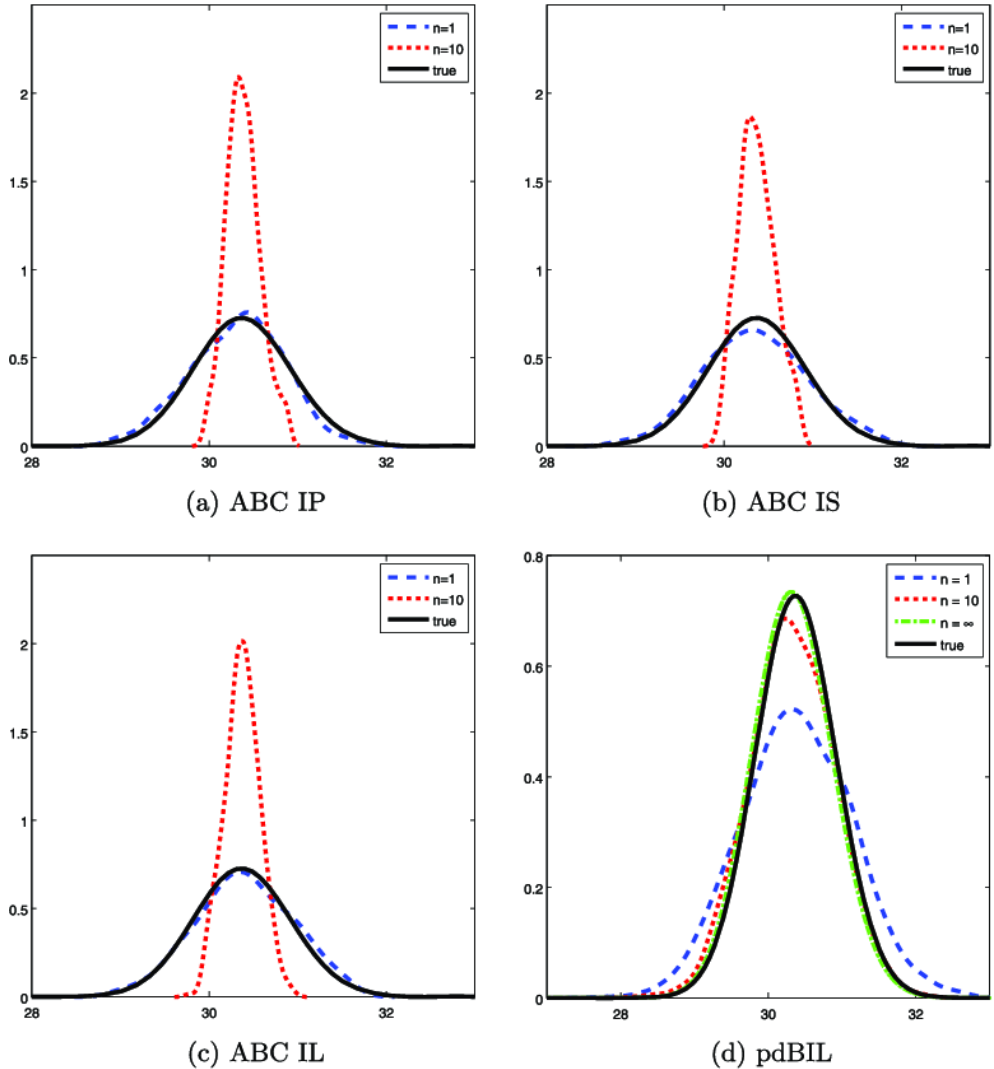}

\caption{Posterior distributions for $\lambda$, the parameter of the
Poisson example for when \textup{(a)} ABC IP, \textup{(b)} ABC IS, \textup{(c)} ABC IL and \textup{(d)}
pdBIL is applied (on-line figure in colour).}\vspace*{-2pt}\label{fig:toy}
\end{figure*}

Here, the data are $N=100$ independent draws from a Poisson
distribution with a mean of $\lambda= 30$, $\vect{y} = (y_1,\ldots
,y_{100}) \stackrel{\mathrm{i.i.d.}}{\sim} \operatorname{Po}(30)$.
The prior is
$\lambda\sim\Gamma(\alpha,\beta)$ (where $\alpha= 30$ and $\beta=
1$), which results in a $\lambda| \vect{y} \sim\Gamma(30+\sum_{i=1}^{100}y_i,101)$ posterior. For such a relatively large value for
the mean of the Poisson distribution, a normal approximation with mean,
$\mu$, and variance, $\tau$, will be reasonable. We use this normal
distribution as the auxiliary model. Here, the auxiliary likelihood,
MLE and score are trivial to compute. The Anderson--Darling test for
normality produced a $p$-value of about 0.576, which indicates no
evidence against the assumption that the normal auxiliary model
provides a good description of the data.

The summary statistic based on this auxiliary model includes the sample
mean, $\bar{y}$, which is a sufficient statistic for the generative
model. Thus, the ABC II approaches can be expected to produce
essentially exact inferences (excluding Monte Carlo error) as long as
the ABC tolerance is low enough. As demonstrated in Figure~\ref{fig:toy},
this is the case. Such sufficiency is not usually achieved
in practice. However, it can be seen that the ABC posterior is grossly
over-precise when the size of the simulated datasets is increased to
1000 (i.e., $n=10$).

In the limit as $n \rightarrow\infty$, the pdBIL method boils down to
a $N(\lambda,\lambda)$ distribution approximating a
$\operatorname{Po}(\lambda)$
distribution. The central limit theorem states that the normal
approximation improves as $\lambda$ increases. Since $\lambda= 30$,
pdBIL can never target the true posterior. The pdBIL target
distribution (for $n \rightarrow\infty$) is proportional to $\lambda
^{\alpha- N/2 - 1}\exp(-(\beta+ N/2)\lambda)\exp(-(2\lambda)^{-1}\sum_{i=1}^Ny_i^2)$ while the true posterior is proportional to $\lambda
^{\alpha+ \sum_{i=1}^Ny_i - 1}\exp(-(\beta+ N)\lambda)$.
Figure~\ref{fig:toy}(d) demonstrates a small amount of bias for the pdBIL
method (this is an illustration of Remark~\ref{rem:BIL_never_correct}).

\begin{figure*}

\includegraphics{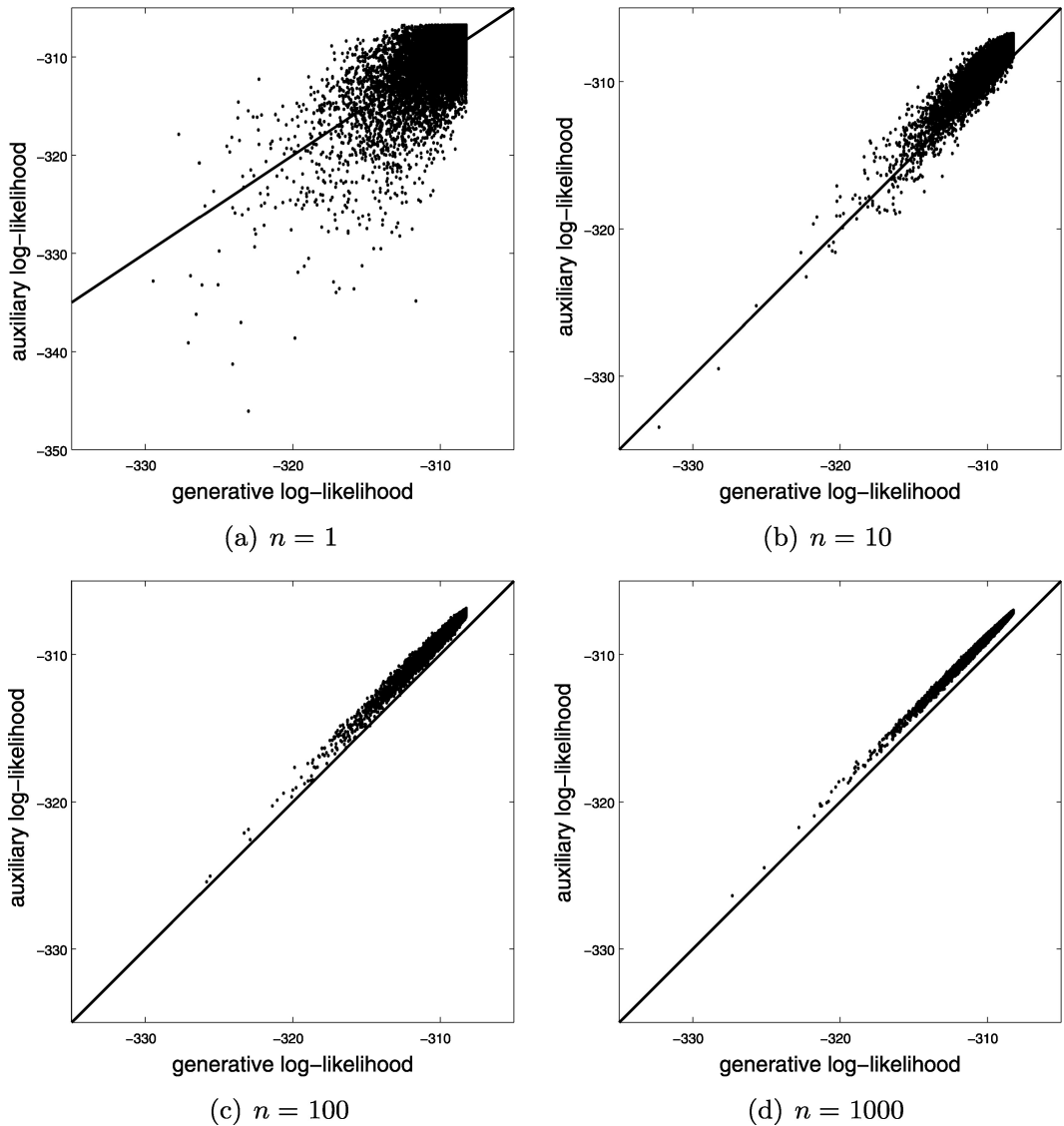}

\caption{Comparison of the generative and auxiliary log-likelihoods for
the toy example calculated during the MCMC pdBIL algorithm with
different values of $n$ [\textup{(a)} $n=1$, \textup{(b)} $n=10$,
 \textup{(c)} $n=100$, \textup{(d)} $n=1000$].}
\label{fig:toy_loglike}
\end{figure*}

\begin{figure*}

\includegraphics{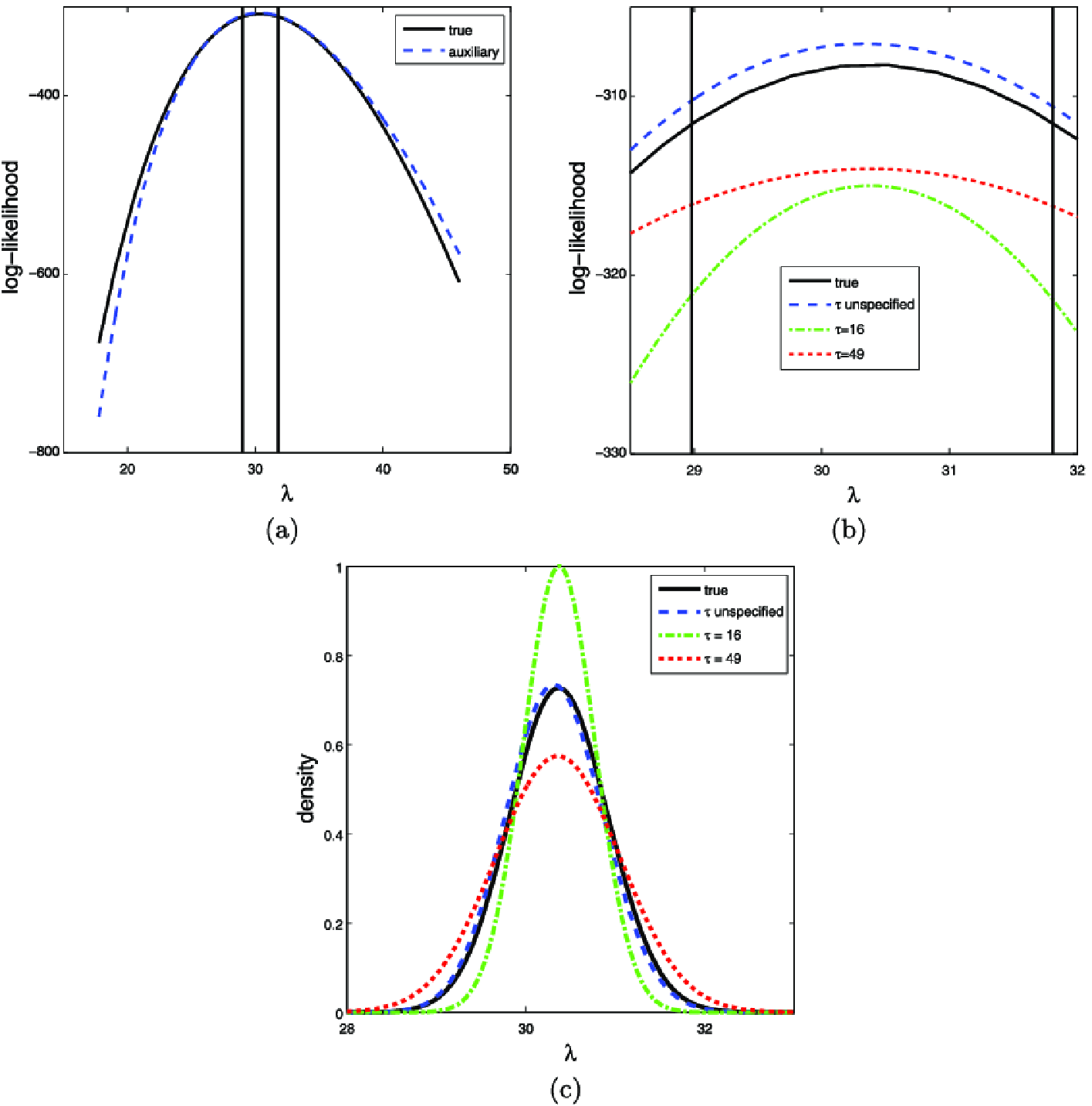}

\caption{Investigating different auxiliary models for the toy example.
\textup{(a)} Comparison of the true and auxiliary log-likelihoods values for
$\lambda$ values within the 99\% highest prior density region. \textup{(b)}
Comparison of true and auxiliary log-likelihoods for different choices
of the auxiliary model. The vertical lines in \textup{(a)} and \textup{(b)} indicate the
bounds of a 99\% credible interval based on the true posterior. \textup{(c)}
Comparison of the posterior distributions for the three different
auxiliary models (on-line figure in colour).}
\label{fig:compare_true_aux}
\end{figure*}

Figure~\ref{fig:toy}(d) presents the results for the pdBIL method
based on simulated dataset sizes of $n=1$ and $n=10$ (results for
$n=100$ and $n=1000$ are even closer to the true posterior but are not
shown on the figure). It is evident from the figure that a more precise
posterior is achieved when using larger simulated datasets, without
necessarily over-shooting the true posterior. Additionally, there was
an increase in the MCMC acceptance rate as $n$ increased. For the $n$
values investigated here, the acceptance rates were roughly 46\%, 67\%,
72\% and 73\% for increasing $n$. These acceptance rates are very high,
especially relative to ABC algorithms which generally suffer from quite
low acceptance probabilities. This example demonstrates that better
inferences using pdBIL can be obtained by increasing the size of the
simulated dataset beyond that of the observed. Unfortunately, ABC
inferences that use a simulated data size larger than that of the
observed data cannot be trusted in the same way (see Remark~\ref
{rem:BIL_increasing_n}).

The reason for improved inferences from pdBIL as $n$ is increased is
apparent from Figure~\ref{fig:toy_loglike}. Here, it can be seen from
increasing $n$ the log-likelihoods of the generative and auxiliary
models are becoming more correlated with the slope of the relationship
becoming approximately one.

Figure~\ref{fig:compare_true_aux}(a) shows the true $\operatorname{Po}(\lambda
)$ and auxiliary $N(\lambda,\lambda)$ log-likelihood values for $\lambda
$ within the 99\% highest prior density region. The vertical lines
indicate the bounds of a 99\% credible interval based on the true
posterior. It can be seen that the auxiliary log-likelihood is a poor
approximation to the true log-likelihood in regions with negligible
posterior support. This is even the case for larger $\lambda$ values
where it would be expected that a normal approximation would be more
appropriate. However, the normal approximation will perform relatively
poorly in the tails of the distribution. It is evident that the
auxiliary likelihood acts as a useful replacement likelihood in the
region of high posterior support [see
Figure~\ref{fig:compare_true_aux}(b)], and this is enough to result in
a good approximation of the true posterior for large $n$.

\begin{figure*}

\includegraphics{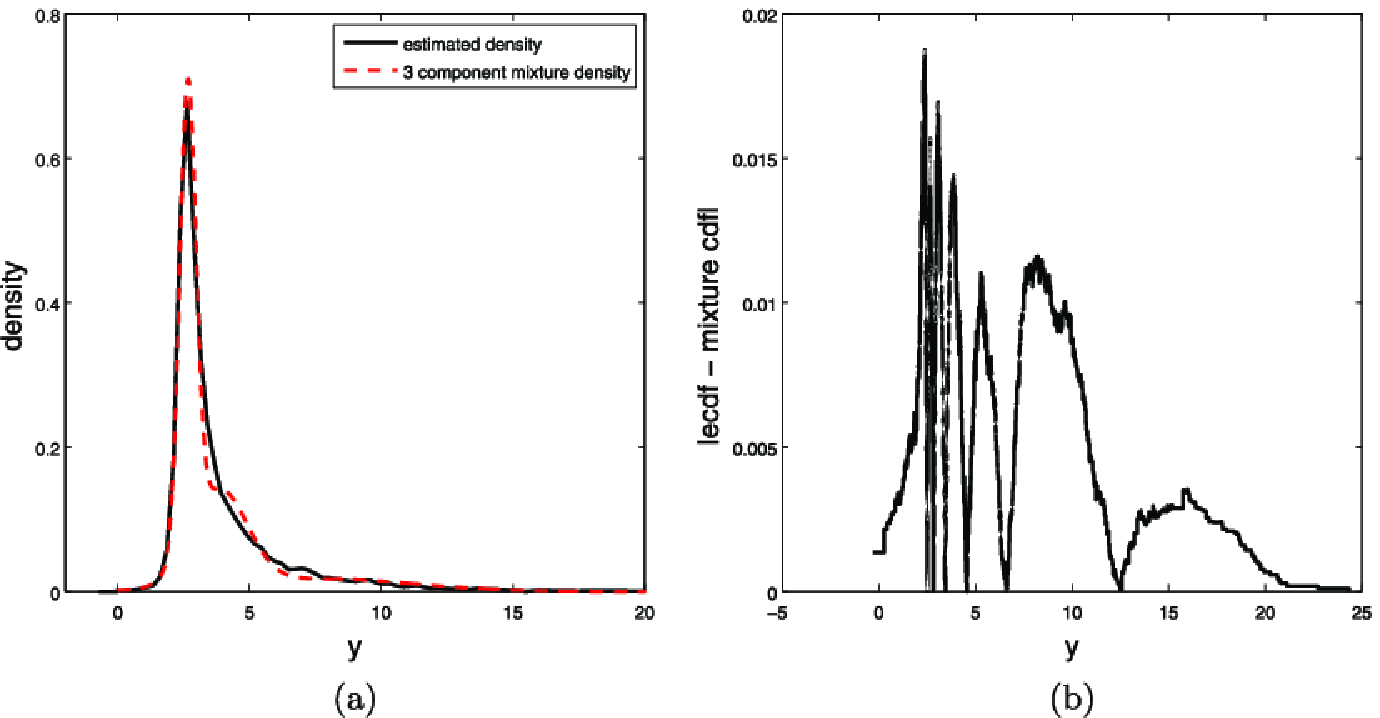}

\caption{\textup{(a)} Nonparametric estimate of the density function based on a
dataset simulated from the $g$-and-$k$ distribution together with the
density of a three component mixture of normals estimated from the
data. \textup{(b)} Absolute value of the difference between the theoretical c.d.f.
of the three component mixture model and the empirical c.d.f. of the data
(on-line figure in colour).}
\label{fig:gandk_data}
\end{figure*}

Finally, we investigate the output from pdBIL when the auxiliary model
is chosen poorly. Figure~\ref{fig:compare_true_aux}(c) shows the
results for when the auxiliary model is $N(\mu,\tau_0)$, where $\tau_0$
is fixed. Here, the pdBIL posterior as $n \rightarrow\infty$ is
proportional to $\lambda^{\alpha-1}\exp(-(\beta- \tau_0^{-1}\sum_{i=1}^Ny_i)\lambda)\exp(-0.5N\tau_0^{-1}\lambda^2)$. Here, we consider
$\tau_0 = 49$ (over-dispersed) and $\tau_0 = 16$ (under-dispersed). The
results show over-precise and conservative results in the
under-dispersed and over-dispersed case, respectively. The
under-dispersed and over-dispersed auxiliary models have thinner and
fatter tails, respectively, than the likelihood of the generative model
in the parameter space well supported by the posterior distribution
(see Figure~\ref{fig:compare_true_aux}). In these cases, the auxiliary
model is not providing a useful replacement likelihood. Just by chance
ABC II is still exact here as $\epsilon\rightarrow0$ since the
parameter estimate for $\mu$ is a sufficient statistic for $\lambda$.

\subsection{$g$-and-$k$ Example} \label{subsec:quantile_example}

\subsubsection{Models and data}

Quantile distributions (or functions) represent a class of
distributions that are defined in terms of their quantile function.
Such functions can be formulated to create more flexible distributions
than other standard distributions. In this example, the focus is on the
$g$-and-$k$ distribution described in, for example, \citet{Rayner2002} (the
reader is also referred to the references therein). This quantile
function, which can also be interpreted as a transformation of a
standard normal random variate, has the following form:
%
\begin{eqnarray}
\label{eq:g-and-k} Q\bigl(z(p);\vectr{\theta}\bigr) &=& a + b \biggl(1 + c
\frac{1-\exp(-gz(p))}{1 + \exp
(-gz(p))} \biggr)
\nonumber
\\[-8pt]
\\[-8pt]
&&\phantom{a +}{}\cdot\bigl(1 + z(p)^2\bigr)^kz(p).\nonumber
\end{eqnarray}
Here, $p$ denotes the quantile of interest while $z(p)$ represents the
quantile function of the standard normal distribution. The model
parameter is $\vectr{\theta} = (a,b,c,g,k)$, though common practice is
to fix $c$ at 0.8, which we do here (see \cite*{Rayner2002}, for a
justification). The likelihood function can be computed numerically,
although this is more expensive than model simulation which is cheaply
implemented for quantile distributions via the inversion method. Full
likelihood-based inference is more expensive than the simulation-based
approaches for the relatively large dataset considered here.

The observed dataset consists of 10,000 independent draws from the
$g$-and-$k$ distribution with $a = 3$, $b = 1$, $c=0.8$, $g = 2$ and $k =
0.5$ (same as considered in \cite{DrovandiPettitt2011}). A
nonparametric estimate of the probability density function based on
these samples is shown in Figure~\ref{fig:gandk_data}. The data exhibit
significant skewness and kurtosis.

We use a three component normal mixture model with 8 parameters as the
auxiliary model. A mixture model is a suitable choice for an auxiliary
distribution since it can be made arbitrarily flexible whilst
maintaining a tractable likelihood function. Therefore, auxiliary MLEs
are computationally easy to obtain [here we use the
Expectation-Maximisation (EM) algorithm] and the subsequent likelihood
can be evaluated cheaply. On the other hand, mixture models can he
highly irregular and the MLE is not consistent in general. The
invariance of the likelihood to a re-labelling of the components causes
an immediate issue for ABC IP, which requires a unique auxiliary
parameter estimate. In an attempt to overcome this, we post-process the
mixture model parameter estimates generated throughout the ABC IP
algorithm by ordering them based on the component means. Since pdBIL
and ABC IL use the likelihood of the auxiliary model, they more
naturally overcome the label switching issue. However, the mixture
model can give other numerical issues such as those resulting from
infinite likelihoods. This would create serious issues for methods that
use the auxiliary likelihood (the auxiliary likelihood would not be
unique). From investigations on the dataset here, it appears that the
likelihood is well behaved and that the modes in the likelihood
correspond only to re-labelling of components. Therefore, we proceed
with ABC IL and pdBIL with caution. The ABC IS method, based on the
score vector, appeared to not have any difficulties accommodating the
auxiliary mixture model.

From Figures~\ref{fig:gandk_data}(a) and~\ref{fig:gandk_data}(b), it can
be seen that there is a correspondence between both the densities and
the cumulative distribution functions of the mixture model and the
data. However, we performed a hypothesis test to assess the
goodness-of-fit of the three component mixture model with a parameter
given by the MLE. The test-statistic was the Kolmogorov--Smirnov
statistic that computes the maximum absolute difference between the
theoretical and empirical c.d.f.s. To avoid any distributional assumption
about this test-statistic, we simulated 10,000 values of this statistic
under the assumption that the mixture model is correct. We found that
the observed test-statistic was exceeded 0.25\% of the time, indicating
strong evidence against the mixture providing a good fit to the data.
Figure~\ref{fig:gandk_data}(b) shows the differences between the
empirical and theoretical c.d.f.'s. However, from Figure~\ref{fig:gandk_data}(a)
it is evident that the mixture model can explain
several features of the true model, and since the dataset size is large
there is a high probability of detecting a difference. Our results
below show that we are able to obtain quite accurate posterior
distributions with the pBII methods despite the lack of fit suggested
by the hypothesis test. In Appendix B of the supplemental article \citep
{DrovandiSupp2014}, we present results from using a four component
mixture model. Unfortunately we found this was substantially more
expensive to apply and resulted in some numerical problems.

\subsubsection{Results}

The proposal distribution in the MCMC for the pdBIL algorithm was
guided using the results in \citet{DrovandiPettitt2011}, who analysed
the same data via a traditional ABC approach that used robust measures
of location, scale, skewness and kurtosis as the summary statistics.

pdBIL was run using $n$ values of 1, 2, 4, 10, 20 and 50 for a number
of iterations given by 1 million, 500,000, 500,000, 200,000, 100,000 and
75,000, respectively. The MCMC acceptance probabilities obtained were
about 2.8\%, 5\%, 7.1\%, 13.1\%, 18.5\% and 20.8\%, respectively. The
average effective sample size (ESS, averaged over the four parameters)
divided by the computing time (in hours) were roughly 63, 127, 106,
124, 70 and 41, respectively. This demonstrates how pdBIL is still
feasible as $n$ increases to a certain point. However, for very large
$n$ the computation becomes unmanageable.

\begin{figure*}[t]

\includegraphics{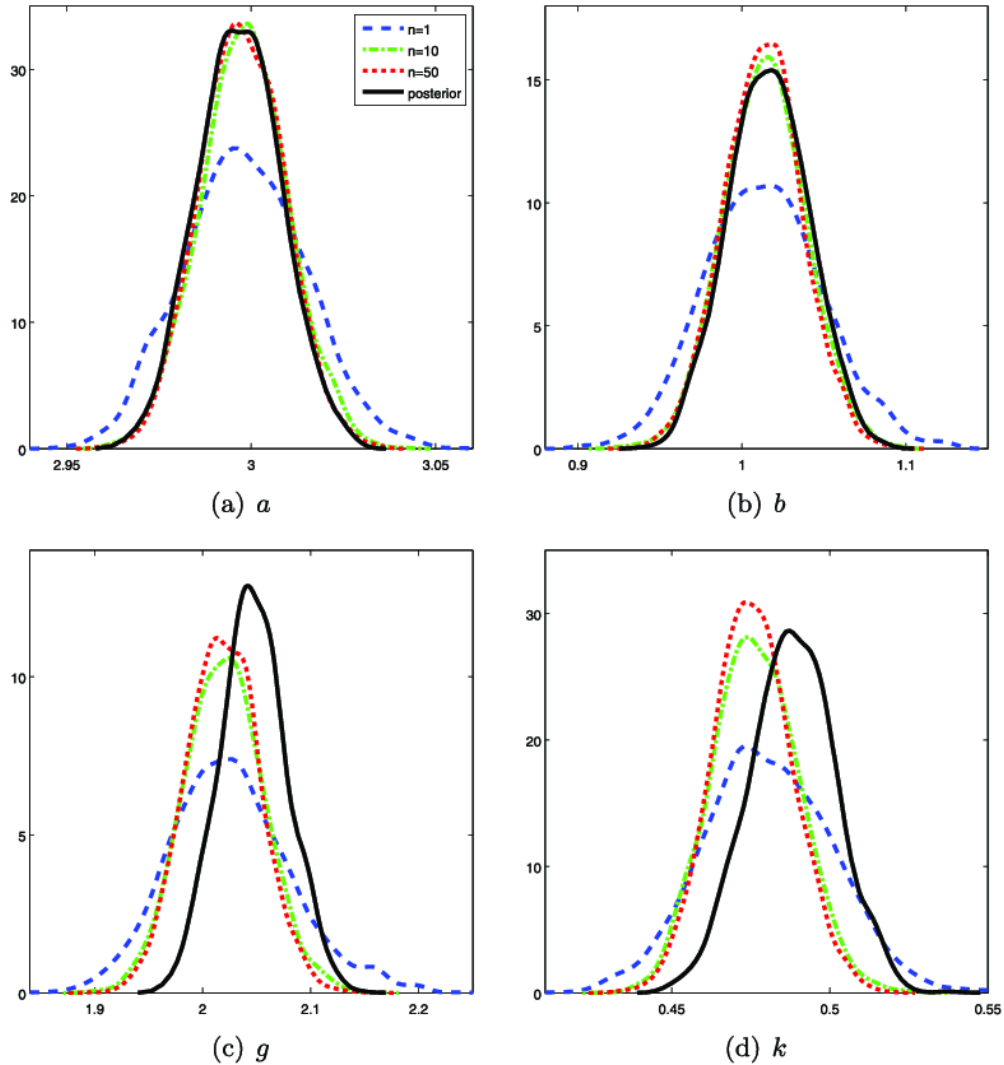}

\caption{Posterior distributions for the parameters of the $g$-and-$k$
model based on the pdBIL approach with $n=1$ (dash), $n=10$
(dot-dash) and $n=50$ (dot). Also shown are results based on
using the true likelihood (solid) (on-line figure in colour).}
\label{fig:gandk_large}
\end{figure*}

Figure~\ref{fig:gandk_large} shows the results for $n=1$, $n=10$ and
$n=50$ (the results for $n=20$ and $n=50$ were quite similar). A very
time consuming exact MCMC algorithm was run for 10,000 iterations to
obtain a gold-standard (producing an average ESS per hour of 6). The
results show an increase in precision of the pdBIL posteriors as $n$
increases. The results for $a$ and $b$ are very accurate, while the
pdBIL posteriors for $g$ and $k$ show some bias (also with a loss of
precision for $g$).

\begin{figure*}[t]

\includegraphics{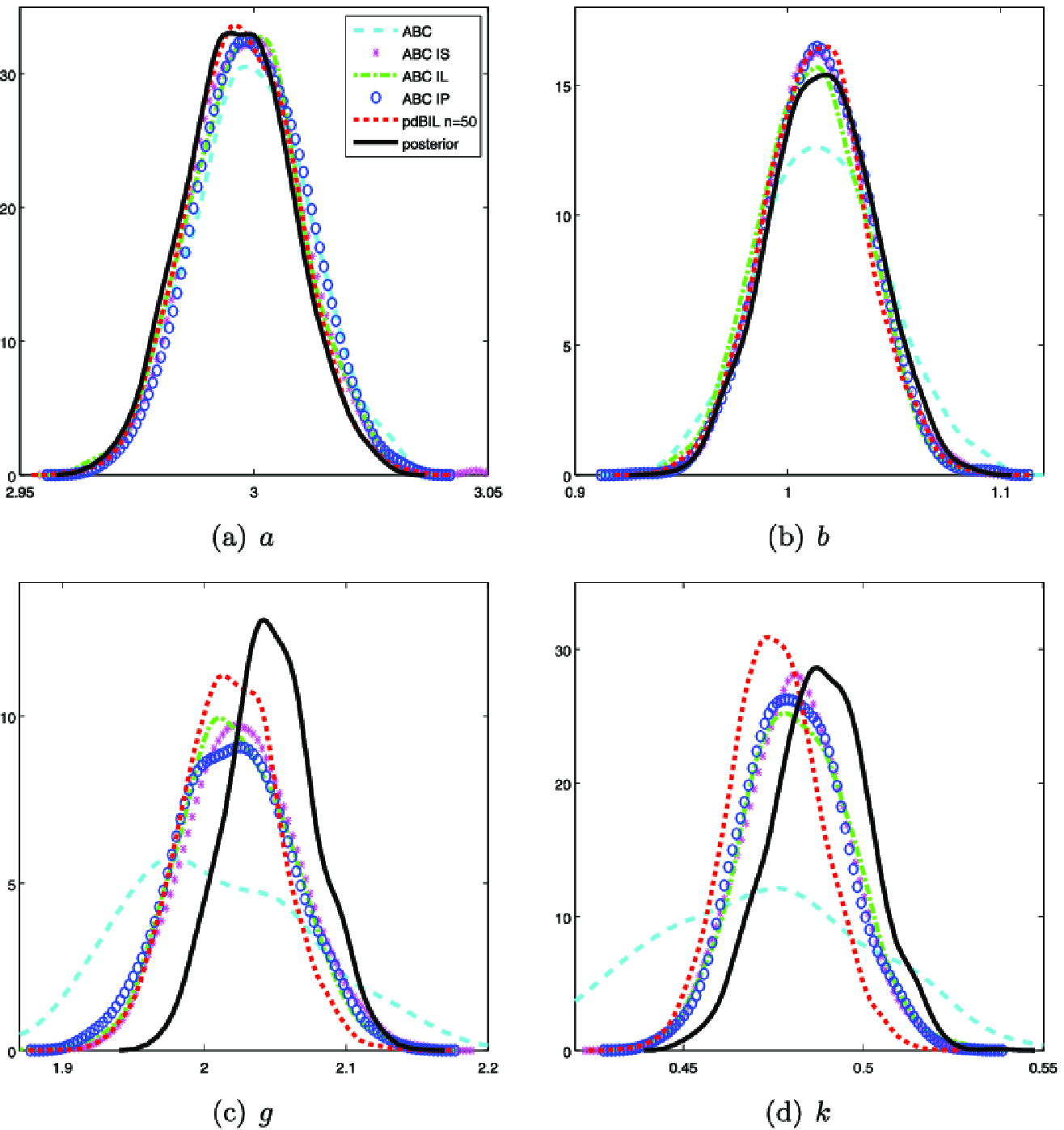}

\caption{Posterior distributions for the parameters of the $g$-and-$k$
model based on the ABC approach of
Drovandi and
Pettitt
(\citeyear{DrovandiPettitt2011}) (dash), ABC IS (star), ABC IL (dot-dash), ABC IP (circle) and
pdBIL with $n=50$ (dot) and results from using the true
likelihood (solid). Note that regression adjustment has been
applied to all ABC results (on-line figure in colour).}
\label{fig:gandk_compare}
\end{figure*}

ABC IP and ABC IL were run for 1 million iterations with a tolerance
tuned to achieve an acceptance rate of about 1\%. Due to the ABC IS
method being so much faster than the other pBII approaches, we aimed
for a relatively lower ABC tolerance and ran the algorithm for more
iterations. More specifically, 7 million iterations were used and the
ABC tolerance chosen resulted in an acceptance rate of 0.3\%. We also
applied a regression adjustment to the (appropriately thinned) ABC II
samples using the approach of \citet{Beaumont2002} in an attempt to
eliminate the effect of the ABC tolerance. In order to apply regression
adjustment for ABC IL, the same post-processing procedure used for ABC
IP was required. Without regression adjustment [see Figure~2 of the
supplemental \mbox{article} \citep{DrovandiSupp2014}], the ABC IS method gave
slightly better results than other ABC II methods, which could be due
to the ability of ABC IS getting to a lower ABC tolerance. The
unadjusted ABC IL results were also slightly better than the unadjusted
ABC IP results. ABC IS produced an average ESS per hour of 90 while the
corresponding number was 50 and 30 for ABC IP and ABC IL, respectively,
showing that the ABC IS method required less time to produce a better
approximation. Regression adjustment offered improvement to all the ABC
II methods. We compared the pBII approaches with the ABC results of
\citet{DrovandiPettitt2011}. It should be noted that we applied a local
regression adjustment to the ABC results here as we found some
improvement for the parameters $a$ and $g$ (results were very similar
for $b$ and $k$ relative to those obtained in \cite
{DrovandiPettitt2011}). The results are shown in
Figure~\ref{fig:gandk_compare}. Overall, the pBII results present a marked
improvement over the ABC analysis of \citet{DrovandiPettitt2011}, with
$g$ seemingly the most difficult parameter to estimate. The ABC II
methods with regression adjustment produced very similar results.
Taking into account accuracy and computational efficiency, ABC IS with
regression adjustment is probably the preferred method. When a four
component auxiliary model was used [Appendix B of the supplemental
article \citep{DrovandiSupp2014}], the ABC II methods with regression
adjustment produced similar results and outperformed pdBIL in terms of
accuracy. Further, the ABC IS approach was able to avoid the heavy
computation associated with fitting the four component mixture model at
every iteration, and thus also avoided other numerical issues such as
the EM algorithm converging to potentially local optima. The ABC II
regression adjustment results showed some improvement for $g$ and $k$
when going from the three to four component mixture model.

\subsection{Macroparasite Example} \label{subsec:cat_example}

\subsubsection{Models and data}

Drovandi, Pettitt and\break
Faddy
(\citeyear{DrovandiEtAl2011}) developed an ABC IP approach to estimate the
parameters of a stochastic model of macroparasite population evolution
developed by \citet{Riley2003} (see also \cite{Michael1997}). Data was
collected independently on 212 cats, who were initially injected with a
certain number of juvenile \textit{Brugia pahangi} parasites. Some time
after each cat was sacrificed, the number of parasites that had reached
maturity were counted and recorded (see \cite{Denham1972}). \citet
{DrovandiEtAl2011} discovered that a beta-binomial model provided a
good description of the data. Below provides a brief review of the
generative and auxiliary models, with the reader referred to \citet
{DrovandiEtAl2011} for more details.

At time $t$, any host is described by three random variables $ \{
M(t),   L(t),   I(t) \}$, where $M(t)$ is the number
of mature parasites, $L(t)$ is the number of larvae and $I(t)$ is a
discrete version of the host's immunity.

It is assumed that each larva matures at a constant rate of $\gamma$
per day. Larvae die at a rate $\mu_L + \beta I(t)$ per larva where $\mu
_L$ represents the rate at which natural death of larvae occurs and
$\beta$ is a rate parameter that describes additional death of larvae
due to the immune response of the host. The acquisition of immunity is
assumed to be dependent only on the number of larvae and occurs at rate
$\nu L(t)$, and a host loses immunity at a rate $\mu_I$ per unit of
immunity. Mature parasites die at a rate of $\mu_M$ adults per day.
Parameters $\gamma$ and $\mu_M$ have been previously estimated at 0.04
\citep{Suswillo1982} and 0.0015 \citep{Michael1997}, respectively.

The data were modelled via a continuous time discrete trivariate Markov
process. Given current values of the states at time $t$, $M(t) = i$,
$L(t) = j$, $M(t) = k$, and a small time increment $\Delta_t$ the
transition probabilities at time $t+\Delta_t$ are given by
%
\begin{eqnarray}
\label{trivariateParasite} %
p(i+1,j-1,k) &=& \gamma j
\Delta_t + o(\Delta_t),
\nonumber
\\
p(i,j-1,k) &=& (\mu_L + \beta k)j\Delta_t + o(
\Delta_t),
\nonumber
\\
p(i-1,j,k) &=& \mu_Mi\Delta_t + o(
\Delta_t),
\\
p(i,j,k+1) &=& \nu j\Delta_t + o(\Delta_t),
\nonumber
\\
p(i,j,k-1) &=& \mu_I k\Delta_t + o(
\Delta_t),
\nonumber
\end{eqnarray}
and the probability of remaining in the same state is one minus the sum
of the above probabilities. Only the final mature count is observed
whilst the immunity and larvae counts are unobserved throughout the
process. Moreover, the immune response variable $I(t)$ is unbounded.
Data generative likelihood-based approaches appear infeasible due to
computational issues (see \cite{DrovandiEtAl2011}). Simulation is
straightforward via the algorithm of \citet{Gillespie1977}. The prior
distributions are: $\nu\sim U(0,1)$, $\mu_I \sim U(0,2)$, $\mu_L \sim
U(0,1)$ and $\beta\sim U(0,2)$.

Here, we denote the observed data as $\vect{y} = (m_1,\ldots,\allowbreak  m_{212})$
where $m_i$ is the mature count for the $i$th host. Covariates for the
$i$th host are given by $l_i$ (initial larvae count) and $t_i$
(sacrifice time).

For the auxiliary model, \citet{DrovandiEtAl2011} capture the
overdispersion via a beta-binomial regression model and take into
account the effect that $t_i$ and $l_i$ have on $m_i$. Denote $\alpha
_i$ and $\beta_i$ as the beta-binomial parameters for the $i$th host.
More specifically, the $i$th observation has the following probability
distribution:
%
\begin{equation}
\label{eq:bb} \quad\  p(m_i | \alpha_i,\beta_i) =
\pmatrix{{l_i}
\cr
{m_i}}\frac{B(m_i + \alpha_i,
l_i - m_i + \beta_i)}{B(\alpha_i,\beta_i)},
\end{equation}
where $B(\cdot,\cdot)$ denotes the beta function. Consider a
re-parameterisation in terms of a proportion, $p_i = \alpha_i/(\alpha_i
+ \beta_i)$, and over-dispersion, $\xi_i = 1/(\alpha_i + \beta_i)$,
parameter. The auxiliary model relates these parameters to the
covariates via
\begin{eqnarray*}
\operatorname{logit}(p_i) &=& f_p(t_i,l_i),
\\
\log(\xi_i) &=& f_{\xi}(t_i,l_i),
\end{eqnarray*}
where
\[
f_{\xi}(t_i,l_i) = f_{\xi}(l_i)
= \cases{ \eta_{100}, &$\mbox{if } l_i\leq100$
\cr
\eta_{200},& $\mbox{if } l_i > 100$ }
\]
and
\begin{eqnarray*}
f_p(t_i,l_i) &=& f_p(t_i)
\\
&=& \beta_0 + \beta_1 \bigl(\log(t_i) -
\overline{\log (t)}\bigr)
\\
&&{}+ \beta_2 \bigl(\log(t_i) -
\overline{\log(t)}\bigr)^2. 
\end{eqnarray*}
Hence, the auxiliary model has the parameter $\vectr{\phi} = (\beta_0,
\beta_1, \beta_2, \eta_{100}, \eta_{200})$ while the generative model
has the parameter $\vectr{\theta} = (\nu,\mu_I,\mu_L,\beta)$.

\begin{figure*}

\includegraphics{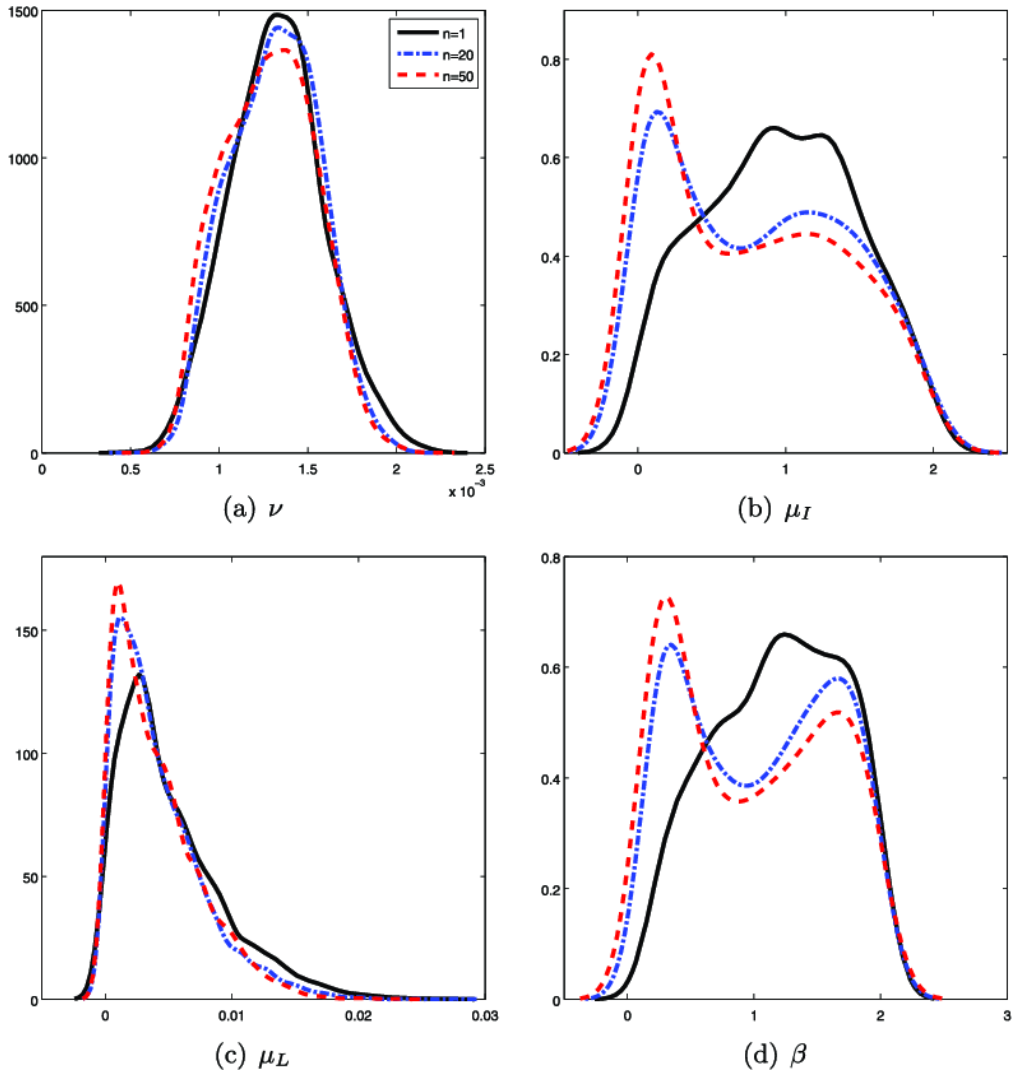}

\caption{Posterior distributions for the parameters [\textup{(a)} $\nu$,
\textup{(b)} $\mu
_I$, \textup{(c)} $\mu_L$, \textup{(d)} $\beta$] of the
macroparasite model based on a
pdBIL approach with $n=1$ (solid), $n=20$ (dot-dash) and $n=50$
(dash) (on-line figure in colour).}
\label{fig:cat_BIL}
\end{figure*}

Using the approach outlined in Appendix C of the supplemental article
(\citeauthor{DrovandiSupp2014},\break \citeyear{DrovandiSupp2014}), we obtained goodness-of-fit $p$-values of 0.37
and 0.47, indicating no evidence against the beta-binomial model
providing a good description of the data. \citet{DrovandiEtAl2011} use
the AIC to select this auxiliary model over competing auxiliary models.

\subsubsection{Results for simulated data}

For validation of the pBII methods for this example, data was simulated
using the same experimental design as the observed data based on the
parameter configuration estimated by \citet{Riley2003}; $\nu=
0.00084$, $\mu_I = 0.31$, $\mu_L = 0.0011$ and $\beta= 1.1$. We found
that the pBII methods were able to recover the parameters $\nu$ and $\mu
_L$ well, $\mu_I$ was determined less precisely and $\beta$ was not
recovered. The data are not particularly informative about $\mu_I$ and
$\beta$ (see \cite{DrovandiEtAl2011}, for more discussion). The ABC IS
gave the most precise posterior distributions for $\nu$ and $\mu_L$ out
of the pBII methods. For full details on the analysis of this simulated
data, see Appendix C of the supplemental article \citep{DrovandiSupp2014}.

\subsubsection{Results for real data}

Here, we used the ABC IP results of \citet{DrovandiPettitt2011} to form
an MCMC proposal distribution. The pdBIL method with $n=1$, $n=20$ and
$n=50$ was run for 1 million, 100,000 and 50,000 iterations,
respectively. Acceptance probabilities of roughly 1.4\%, 23.5\% and
28.2\%, respectively, were obtained. The average ESS per hour was 37,
79 and 58, respectively. The substantial increase in acceptance
probability allowed us to use fewer iterations. The results are shown
in Figure~\ref{fig:cat_BIL}. The figures suggest that we are not able
to gain any additional information from the data for the parameter $\nu
$ from the pdBIL approach by increasing $n$. However, an increase in
precision is obtained for $\mu_L$ as $n$ is increased. The posteriors
are shifted slightly for the other two parameters, however, they are
still largely uninformative, although the posterior for $\mu_I$ for
large $n$ may indicate some preference for smaller values of $\mu_I$.

\begin{figure*}[t]

\includegraphics{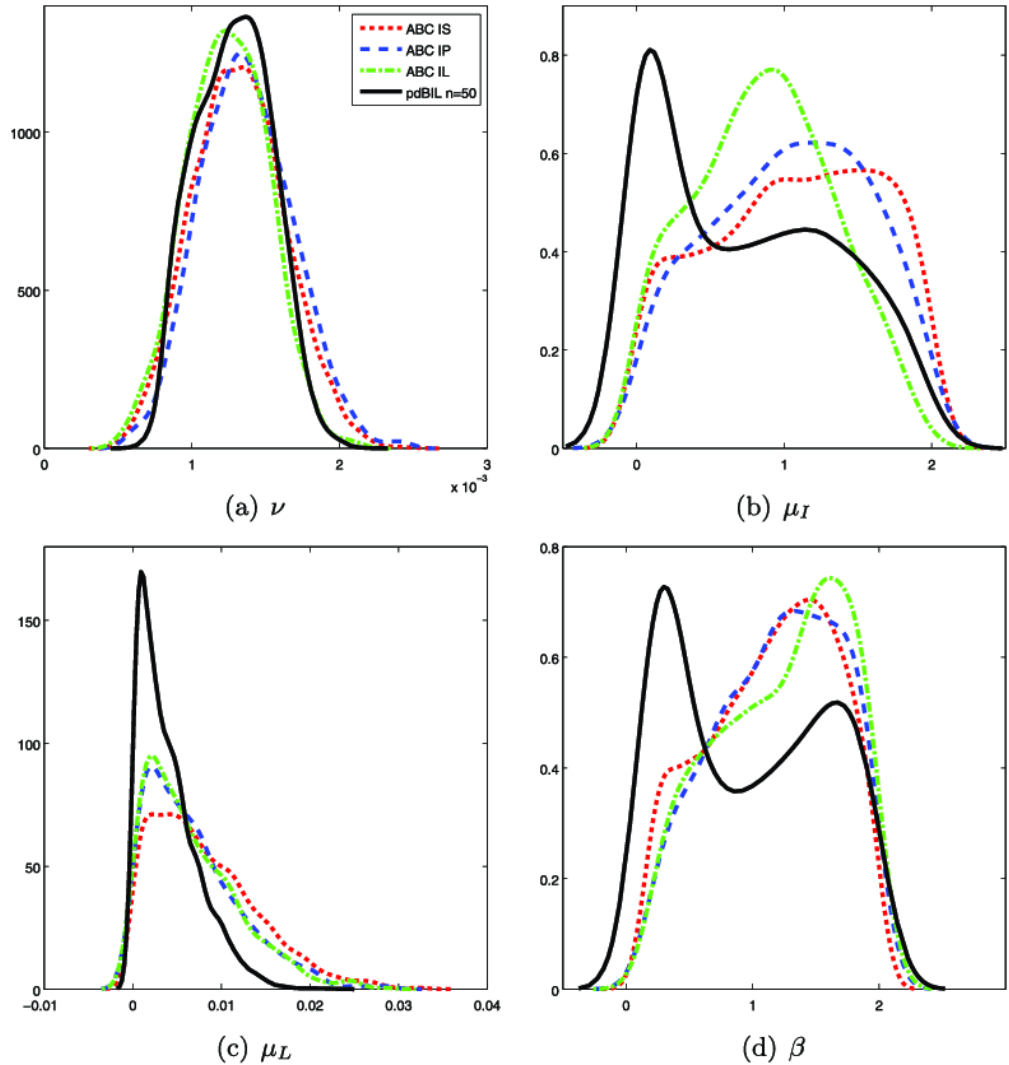}

\caption{Posterior distributions for the parameters [\textup{(a)} $\nu$, \textup{(b)} $\mu
_I$, \textup{(c)} $\mu_L$, \textup{(d)} $\beta$] of the macroparasite model based on ABC
IP (dash), \mbox{ABC IS} (dot), ABC IL (dot-dash) and pdBIL
(solid) (on-line figure in colour).}
\label{fig:cat_ii_compare}
\end{figure*}

We now compare the results of pdBIL with ABC. ABC IP and ABC IL MCMC
algorithms were all run for 1 million iterations. The ABC IP and ABC IL
tolerances were chosen so that the acceptance rate was about 1.5\%. Due
to the increased computational efficiency of ABC IS, we ran this
algorithm for 20 million iterations and tuned the tolerance to obtain
an acceptance rate of about 0.1\%. ABC IP and ABC IL used about 15
hours of computing time while ABC IS only required 11 hours even though
20 times more iterations were run.

\begin{figure*}

\includegraphics{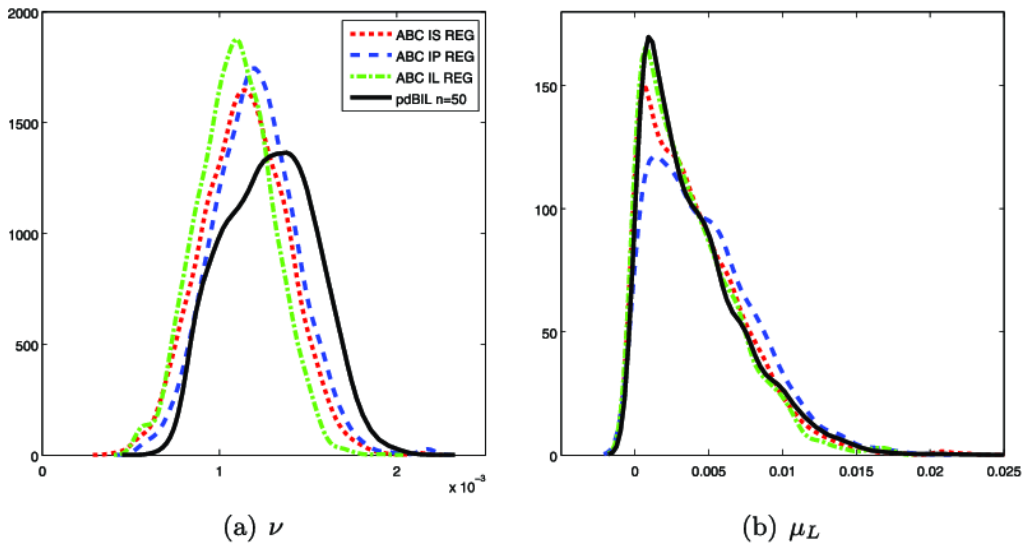}

\label{figsub:cat_ii_compare_nu_reg}
\label{figsub:cat_ii_compare_muL_reg}
\caption{Posterior distributions for the parameters [\textup{(a)} $\nu$ and \textup{(b)}
$\mu_L$] of the macroparasite model based on ABC IP (dash), ABC IS
(dot), ABC IL (dot-dash) and pdBIL (solid). Regression
adjustment has been performed on the output of the ABC II approaches
(on-line figure in colour).}
\label{fig:cat_ii_compare_reg}
\end{figure*}

The estimated posterior densities (after appropriate thinning) for the
different approaches are presented in Figure~\ref{fig:cat_ii_compare}.
In general, the data are not informative about the $\mu_I$ and $\beta$
parameters, so we turn our focus to the parameters $\nu$ and $\mu_L$.
We note that it is difficult to compare the approximations without
having available a gold standard. It can be seen that pdBIL produces
the most precise inferences for the parameters $\nu$ and $\mu_L$.
Despite being able to reduce the ABC tolerance, the ABC IS method
appears to be the least precise. This is in contrast to the results for
the simulated data in Appendix C of the supplemental article \citep
{DrovandiSupp2014}, where ABC IS produced the most precise results.

Regression adjustment was also applied to the ABC II methods in an
attempt to reduce the effect of the ABC tolerance. These adjustments
were applied individually to $-\log(\nu)$ and $\sqrt{\mu_L}$ [see
Appendix C of the supplemental article (Drovandi, Pettitt and Lee, \citeyear{DrovandiSupp2014})] and the
results are shown in Figure~\ref{fig:cat_ii_compare_reg}. The
regression adjustment does increase the precision of the ABC II
posteriors. The regression adjustment appears to shift the modes of the
ABC II results slightly for $\nu$. For $\mu_L$, the regression
adjustment brings the ABC II results closer to that obtained by pdBIL
for $n=50$.

\section{Discussion} \label{sec:Discussion}

This paper has provided an extensive comparison of pBII methods, from
theoretical, practical and empirical perspectives. We discovered that
the pdBIL method of \citet{Gallant2009} is fundamentally different to
ABC II approaches developed in the literature. More specifically, we
showed that pdBIL can produce better approximations by increasing the
size of the simulated datasets as long as the auxiliary model provides
a useful replacement likelihood for the generative likelihood for a
variety of $\vectr{\theta}$ values. In contrast, ABC methods (including
those that use II to form the summary statistic) should simulate
datasets the same size as the observed. The pdBIL method has the
additional advantage of not having to determine an appropriate ABC
tolerance. Furthermore, we found that increasing the size of the
simulated dataset beyond that of the observed does not necessarily make
computation infeasible due to the increase in statistical efficiency.
However, it is of interest to determine the size of the simulated
dataset upon which negligible improvement will be obtained. This
requires further research.

We have also established that BIL is a rather flexible framework since
the synthetic likelihood approach of \citet{Wood2010} is a pBIL method
that applies a parametric auxiliary likelihood to the summary statistic
likelihood while ABC can be recovered by selecting a specific
nonparametric auxiliary model. Our focus in this paper has been on the
pBIL method where a parametric auxiliary model is proposed for the full
data likelihood. However, the ideas in this paper may carry over to
when the auxiliary model is applied to a summary statistic likelihood
as in \citet{Wood2010}.

For the pBIL method to have some chance of a good approximation to the
true posterior for the specified generative model, it is important that
the auxiliary model is able to well fit data simulated from the
generative model for parameter values within nonnegligible posterior
regions, at least in the majority of simulations. It would be possible
to perform a goodness-of-fit test on the auxiliary model for every
dataset generated from the proposed model during the MCMC pBIL
algorithm in order to assess the usefulness of the auxiliary model in
the context of the pBIL method. This is the subject of further research.

In this paper, we have not addressed the issue of which ABC II method
provides the best approximation. ABC IS is much faster (when the
auxiliary score vector is analytic) and requires only weak assumptions,
but did not always outperform the other ABC II methods in the examples
considered in this paper. The ABC IP and ABC IL methods differ only in
their discrepancy function and it is not clear if one discrepancy
function dominates the other across applications. Furthermore, it
remains unknown if the auxiliary parameter estimate or auxiliary score
carries the most information in the observed data. It could be that the
optimal choice of ABC II approach is problem dependent. Until further
research is conducted, we suggest trying all three methods (assuming
that ABC IP and ABC IL are computationally feasible). One approach to
speed up ABC IP and ABC IL might be to start with a computationally
simple but consistent estimator (e.g. the method of moments) and apply
one iteration of a Newton--Raphson method to produce an asymptotically
efficient estimator (\cite{Cox1979}, page~308) in a timely manner.

From a practical perspective, these methods have led to improved
approximate analyses for two substantive problems compared with that
obtained in \citet{DrovandiEtAl2011} and \citet{DrovandiPettitt2011}.
Across applications considered in this paper, ABC IS was the most
computationally efficient and led to good posterior approximations.

Overall, pdBIL avoids having to choose an ABC discrepancy function and
the ABC tolerance. If an auxiliary model can be proposed that satisfies
a rather strong condition, more precise inferences can be obtained by
taking $n$ large, which we showed is still computationally feasible
with MCMC pdBIL up to a point. However, ABC II appears to provide a
more general framework for pBII problems, due to the extra flexibility
of being able to incorporate additional summary statistics outside the
set formed from the auxiliary model and potentially providing better
approximations when the auxiliary model is a simplified version of the
generative model. It is this extra flexibility that may see ABC II as
the method of choice as ever-increasingly complex applications are encountered.

\section*{Acknowledgements}

The first two authors were supported by the Australian Research Council
Discovery Project\break DP110100159. The authors would like to express their
gratitude to the referees and editorial team associated with this
paper. The comments and suggestions provided have led to a
substantially improved paper. The authors would like to thank Edwin
Michael and David Denham for access to the macroparasite data. The
first author is grateful to Ewan Cameron, Richard Everitt, Dennis
Prangle, Andy Wood and Christian Robert for useful discussions about
this work. The first author is also grateful to Warwick University
(where this work was partially completed) for providing some funding
for a three month visit.

\begin{supplement}[id=suppA]
\stitle{Supplement to ``Bayesian Indirect Inference Using a~Parametric Auxiliary Model''\newline}
\slink[doi,text={10.1214/14-STS498SUPP}]{10.1214/14-STS498SUPP} 
\sdatatype{.pdf}
\sfilename{sts498\_supp.pdf}
\sdescription{This material contains a simple example to supplement
Section~\ref{s3.1} and additional information and results to supplement
the examples in Sections~\ref{subsec:quantile_example} and \ref{subsec:cat_example}.}
\end{supplement}

%

\end{document}